\documentclass[11pt]{article}
\usepackage{amsmath,amsfonts,amssymb}
\usepackage{latexsym}
\usepackage{dcolumn}
\usepackage{graphicx,epsfig}
\usepackage{amsthm}
\usepackage{color}
\usepackage{graphicx}
\usepackage{caption}
\usepackage{rotating}
\usepackage{tikz}
\usepackage{float}
\begin{document}
\setcounter{page}{1}

\pagestyle{plain}
\pagestyle{plain} \vspace{1cm}
\begin{center}
\Large{\bf Rupture process of 2013 Shonbe earthquake sequence}\\
\small \vspace{1cm} {\bf P. Dehghani$^{\,a,}$\footnote{parham.dehghani@emu.edu.tr}}, \quad {\bf M. Pakzad$^{\,b,}$\footnote{pakzad@ut.ac.ir}}, and \quad {\bf M. R. Gheytanchi$^{\,b,}$\footnote{mrghchee@ut.ac.ir}}\\
\vspace{0.25cm}
$^{a}$Department of Physics, Faculty of Arts and sciences,\\
Eastern Mediterranean University,\\
Famagusta, Northern Cyprus via Mersin 10\\

\vspace {0.35cm}

$^{b}$ Institute of Geophysics, University of Tehran,\\
Tehran, Iran\\
\end{center}

\vspace{1cm}
\begin{abstract}
 In this study, we have used point-source approximation with the premise that it is compatible with the rupture process of the predominant faulting geometry caused by the initiation of Shonbe seismic sequence. First, all recorded earthquakes during six months after the main shock, 2013 April 9th, in the vicinity of khaki anticline near Shonbe, are relocated with the probabilistic nonlinear method based on Bayesian inference. The synthetic test is done, with the exact distribution of real stations on hand. Thus, the proper value for the parameter that can explicitly affect the error results is calculated to be 0.01. After performing relocation task, 98 relocated events with horizontal errors less than 5 km from all 373 recorded events are taken into account for further analysis of the spatial characteristics of the sequence. In advance, point source inversion is implemented for 21 events greater than 4.5 Mn. Then, supplementary investigations and stability tests are performed for the results of the main shock. Accordingly, we are convinced that the gathered results are reliable enough to be used for the seismotectonic analysis of the region resulting from the sequence. Eventually, the reverse slip vector with substantial left-lateral strike slip component (in some cases) is modeled employing the distribution of inverted centroids and resulting faulting geometry for the chosen events. A southward dip of the probable causative fault is seen, furthermore depth range of the rupture is localized within 8 to 12 km, in the lowermost part of the sedimentary cover.\\
{\bf Keywords}: Point-source inversion, Bayesian inference, Nonlinear relocation, Shonbe, Seismotectonics
\end{abstract}

\section{Introduction}
On April 9th 2013, an earthquake of 6.1 Mn and 6.3 Mw hit Dashti, more precisely Shonbe, based on the report of IGUT (Institute of Geophysics of University of Tehran). Epicenter was reported at 16 km far from Kaki, 27 km far from Khormoj, and 90 km far from Bushehr. Although epicenter was close to Bushehr nuclear plant, this critical unit remained undamaged based on the investigations afterwards. The most damages were revealed in small city of Shonbe and 23 near villages. A rough estimation revealed that almost 10000 people over the region were affected by this event. Moreover, a 5.2 Mn earthquake (12:29 local time, April 10th) struck Dashti again and damaged many people and facilities over the region. More precisely, this seismic sequence following 6.1 Mn main shock included 372 aftershocks greater than 3.0 Mn. The phase information was recorded from the resources including IGUT, IIEES, and some Arabic stations over 6 months after the main shock.

In this study, we precisely explore this seismic sequence. First, relocation of the events is performed using Nonlinear Bayesian method. Then, seismic point source inversion of the main shock is acquired afterwards. Furthermore, a comprehensive discussion on the stability of the results is presented. Finally, seismotectonics induced by the sequence is interpreted then.

\section{The Region of study}

Area of interest is located in Zagros seismotectonic province (Alavi 1994). It is limited to the Mountain Front Fault (MFF) to the northeast and to the Zagros Foredeep Fault (ZFF) to the southwest. Zagros seismotectonic province includes different morphotectonic regions (Berberian 1976a; Berberian 1976b; Berberian 1976c; Berberian 1981; Berberian 1995) from which Zagros Foredeep region is relevant to the area of interest in this study. Some of the important fault systems affecting the tectonics of the region are as below (Baker et al. 1993):

\textbf{Mountain Front Fault (MFF).} This fault is a discontinuity between Zagros Simply Folded Belt (ZSFB) and High Zagros (HZ). It is considered as one of the important hidden fault systems over the region with northward dip and dominant reverse slip.

\textbf{Zagros Foredeep Fault (ZFF).} This fault is located parallel to MFF at the height of about 100 meters from the sea level and detaches Zagros Foredeep region from Zagros Coastal Plain.

\textbf{Kazerun Fault.} This fault is 450 km length with dominant right-lateral strike slip mechanism and lies on north-south direction. It is located within the seismogenic part of whole Zagros seismotectonic province and can be distinguished 15 km far away from Kazerun city.

\textbf{Borazjan Fault.} This fault is located southwest of Kazerun Fault with 180 km overall length. Its induced mechanism of the tectonics is reverse-slip. Consequently, most of the strike-slip tectonic activities are absorbed by Kazerun fault system.

\textbf{Karehbas Fault.} This fault is located 90 km to the east of Bushehr area and has about 130 km length. Also its tectonic mechanism is reverse, like most of the faults over this region. More than 109 seismic events with magnitude greater than 5.0 occurred within the region of study over 100 years ago. This explicitly shows the high rate of seismicity over this region (Ambraseys and Melville 1982). Furthermore, inspections revealed more than 10 deadly events that happened from the years 1500 to 2000.
\begin{figure}[ht!]
\centering
\includegraphics[width=9cm]{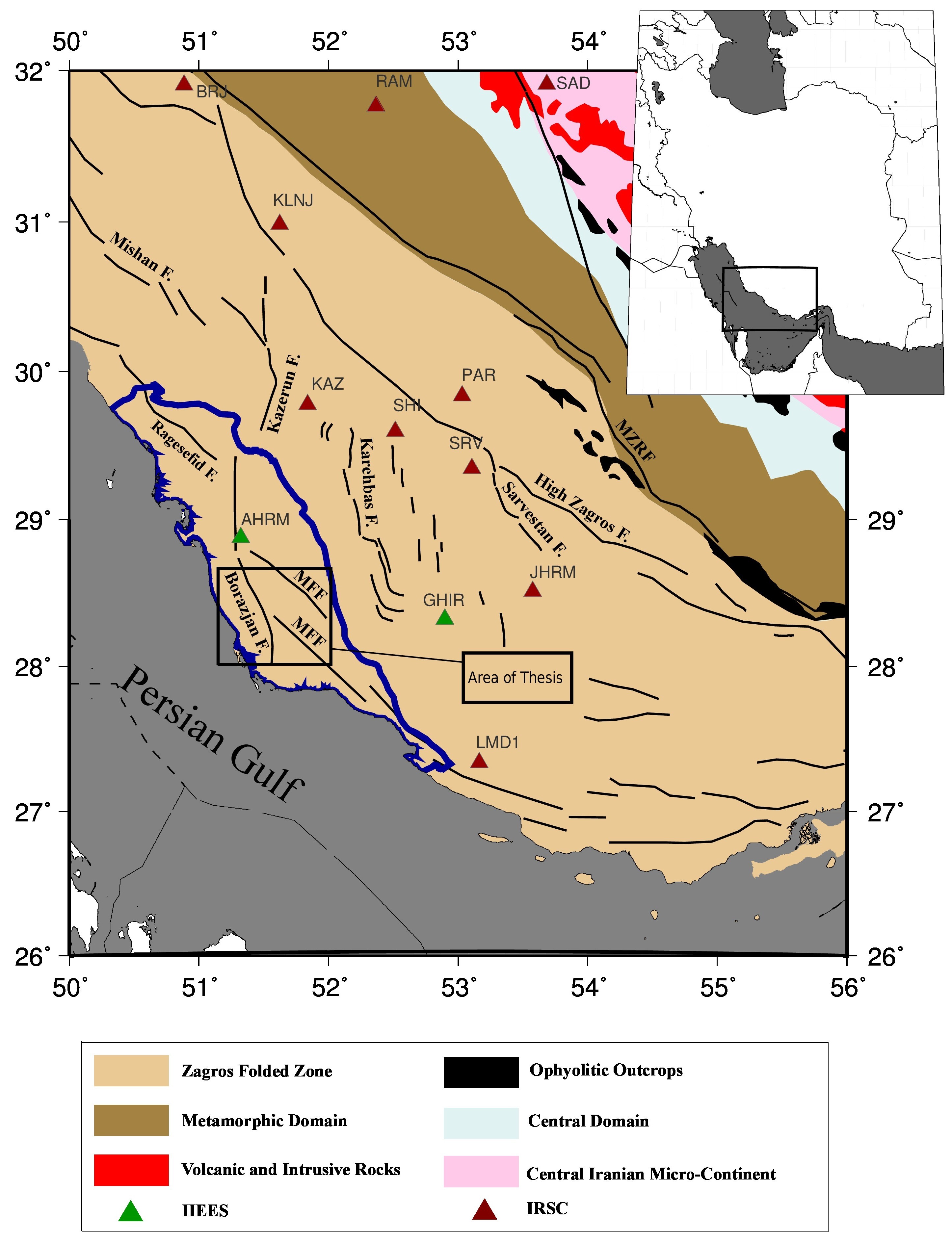}
\caption{\small {Area of study including IIEES and IRSC seismic stations.}}
\end{figure}

\section{Relocation of the seismic sequence}

In this study, we first attempt the relocation of 373 seismic events greater than 3.0 Mn over the region located within latitudes 27.5 to 29.5 N and longitudes 50.5 to 52.5 E, with the use of nonlinear probabilistic method (Geiger 1912; Tarantola and Valette 1982; Lomax et al. 2000, Lomax and Curtis 2001; Lomax 2005). To improve recorded phases and diminish azimuthal gap, we have employed data recorded in IGUT, IIEES, and some Arabian stations. Because of the far distances between most of the stations and the area of the study, we use Sg, Sn, Pg, and Pn phase data. The used three-dimensional grid for relocation task is centered at 28.57 N latitude and 51.97 E longitude and contains all the stations with 600 km distance eastward, westward, northward, and southward from the center. This three-dimensional grid features 130 km depth from the ground to compute the velocity model and consequent travel times to each station. In the figure 2, all the stations of IIEES, IGUT, and the Arabian ones within this grid can be seen. Phase data of these stations are exploited to perform nonlinear relocation task. In this section, 373 seismic events of Shonbe sequence are relocated employing optimized values of the affecting parameters, for instance LocGau2 parameter in the NLLOC software.
\begin{figure}[ht!]
\centering
\includegraphics[width=9cm]{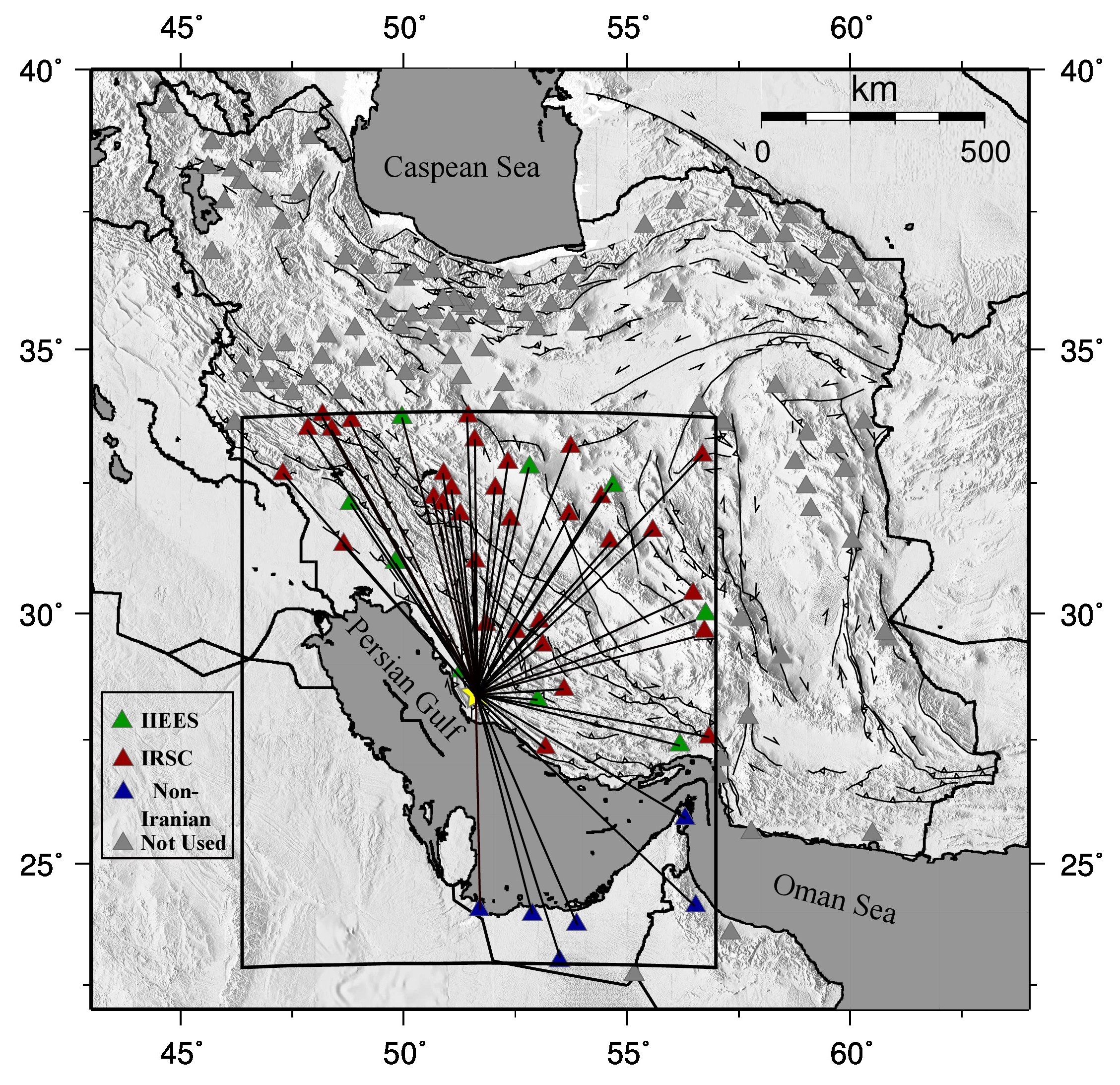}
\caption{\small {Used seismic stations to build travel time grid.}}
\end{figure}
All the events are gathered during 6 months after the main shock, and their phase data are retrieved from IRSC, IIEES, and some Arabian stations. As mentioned before, NLLOC software is employed using EDT probability function that diminishes the unwanted effect of untrue arrival times. Moreover, we choose events to be relocated with at least 15 phase data including S and P phases. Regarding sampling method, Oct-tree (Lomax et al. 2000) is employed because of its pace and accuracy to seek global maxima versus local ones. Resulting depth and epicentral errors, retrieved from covariance matrix of sampled points, are shown in the figure 6 (A).

\subsection{Optimization of the parameters}

One of the present methods to estimate real errors in an inverse problem is to use forward modelling with known data and then take back to inverse problem (here relocation of seismic events) to find the solution and real errors. Importance of this approach becomes higher when all the features or variables affecting the solution of a forward problem cannot be theorized and expressed with mathematics. More important issue is to seek a realistic complicated three-dimensional earth model that can yield forward solutions comparable with real data. To be realistic, this complex model of the earth is not accessible, so an approach has to be proposed to estimate real errors. Importance of real error estimation increases when we face the inverse problem of seismic event depth. With reliable depth estimation on hand, rupture analysis and the resulting seismogenic zone can be analyzed with higher certainty.

In this section, we seek optimized value of LocGau2 parameter by artificial hypocenter inversion of the known events using all the stations shown in the figure 2. This parameter considerably affects the result of inversion and the errors, then more careful choice of its value can improve the result of relocation process. Furthermore, a Gaussian error with standard deviation of 0.1 second is considered as Pg and Pn phase picking error.

To build artificial events, 5 vertical profiles of events with 20 nodes each ranging from 1 km to 39 km depth with 2 km vertical distance are considered. The central profile is put in the center of relocation grid, and other 4 profiles are placed 4 km distant in 4 directions (east, west, north and south) from the center. We employ Time2EQ package of NonLinLoc software (Lomax et al. 2000) for the calculation of travel times. The one-dimensional velocity model of Hatzfeld (Hatzfeld et al. 2003), retrieved from local events for central Zagros, is exploited. Furthermore, 42 stations, exactly the ones used for the main shock, are used to build travel times of Pg and Pn phases considering 0.1 s Gaussian error of phase picking.

After the calculation of travel times for all events and stations, inverse problem of finding hypocenters of all events is performed, while different values of LocGau2 parameter ranging from 0.005 to 0.05 are examined. Finally, to inject the pivotal effect of untrue velocity model of the earth, new velocity model used in IGUT is employed to perform inversion task.
\begin{figure}[ht!]
\centering
\includegraphics[width=6cm]{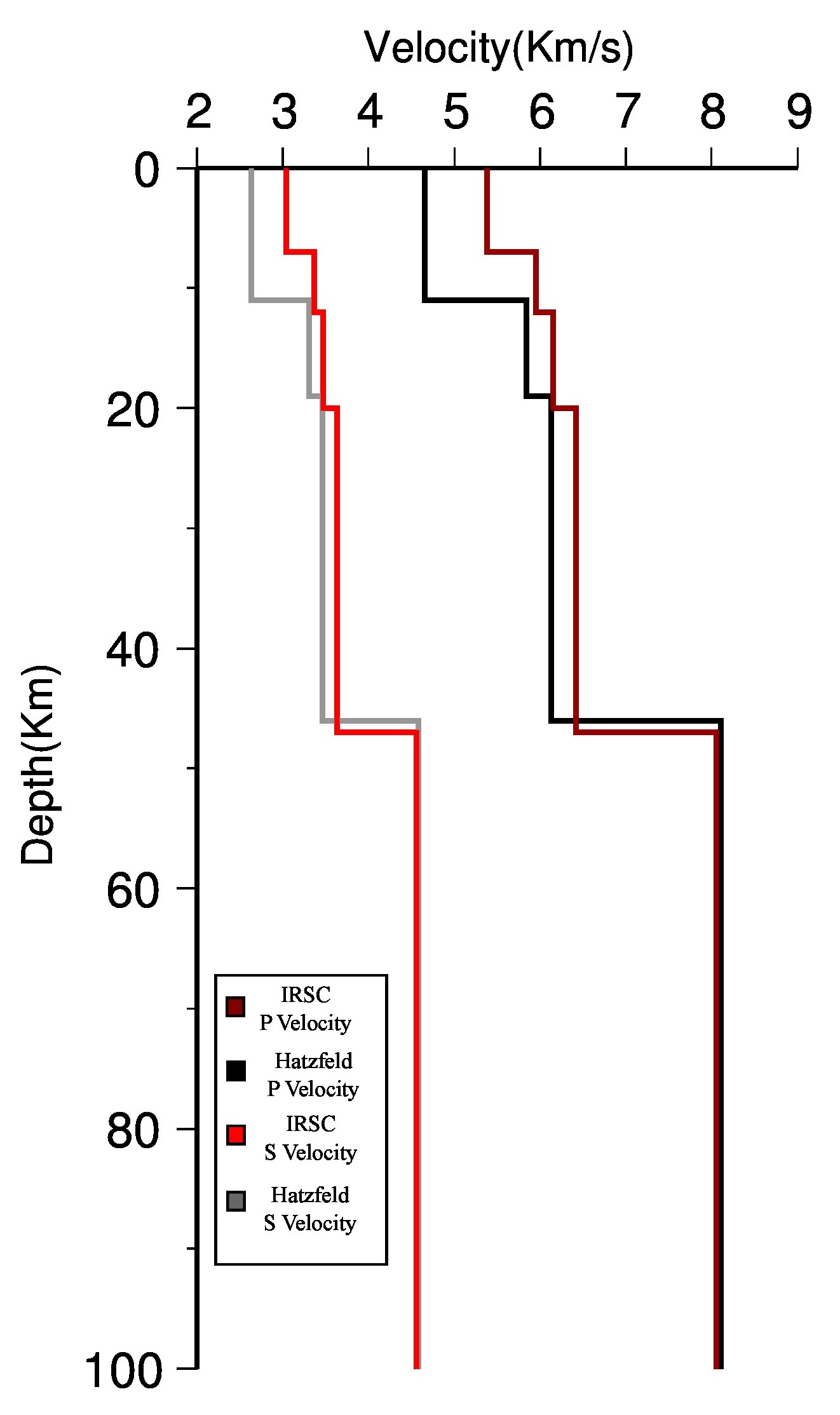}
\caption{\small {Hatzfeld and IRSC one-dimensional velocity model in comparison.}}
\end{figure}

Real versus modeled errors are compared with respect to each LocGau2 parameter shown in the histogram for all artificial events, as seen in figure 4. In this figure, both real and modeled errors of epicenter and depth are shown for all 100 artificial events for each value of LocGau2 parameter. Modeled errors deviate from real errors considerably as LocGau2 parameter increases, so that the big value of this parameter cannot truly lead to the modeled errors close to the real ones.

\begin{sidewaysfigure}
    \centering
    \includegraphics[width=6.5cm]{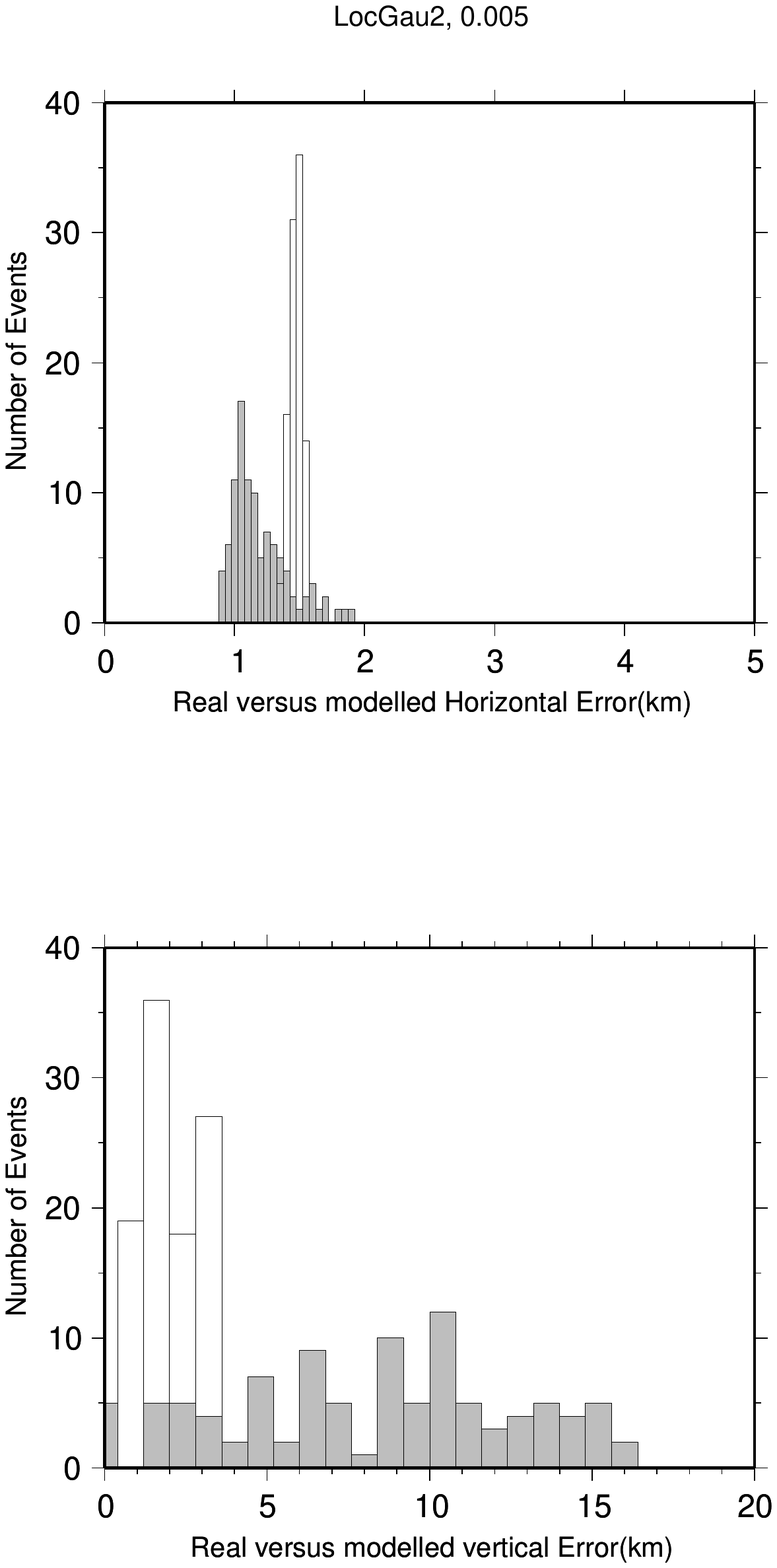}
\includegraphics[width=6.5cm]{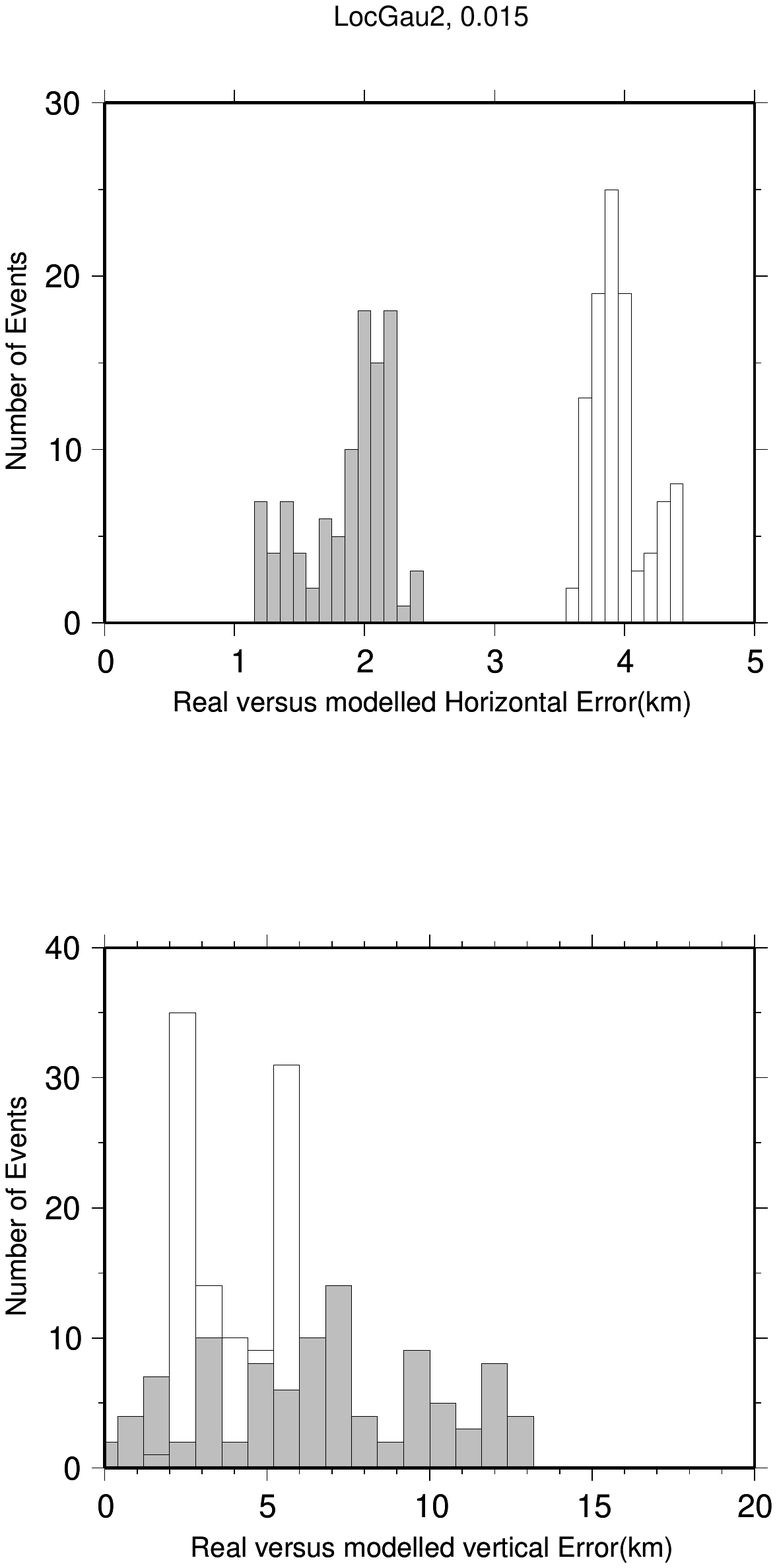}
\includegraphics[width=6.5cm]{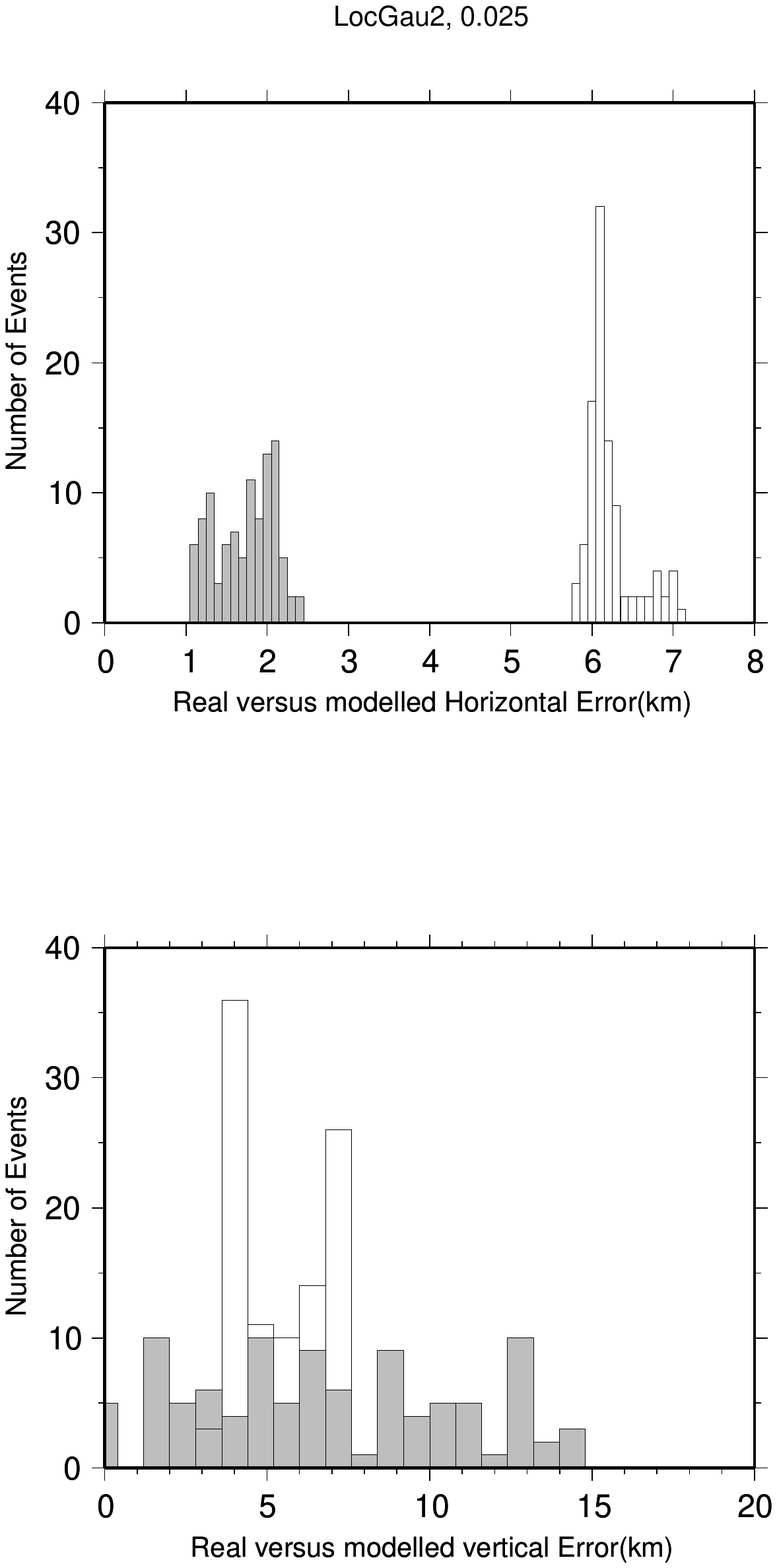}
\includegraphics[width=6.5cm]{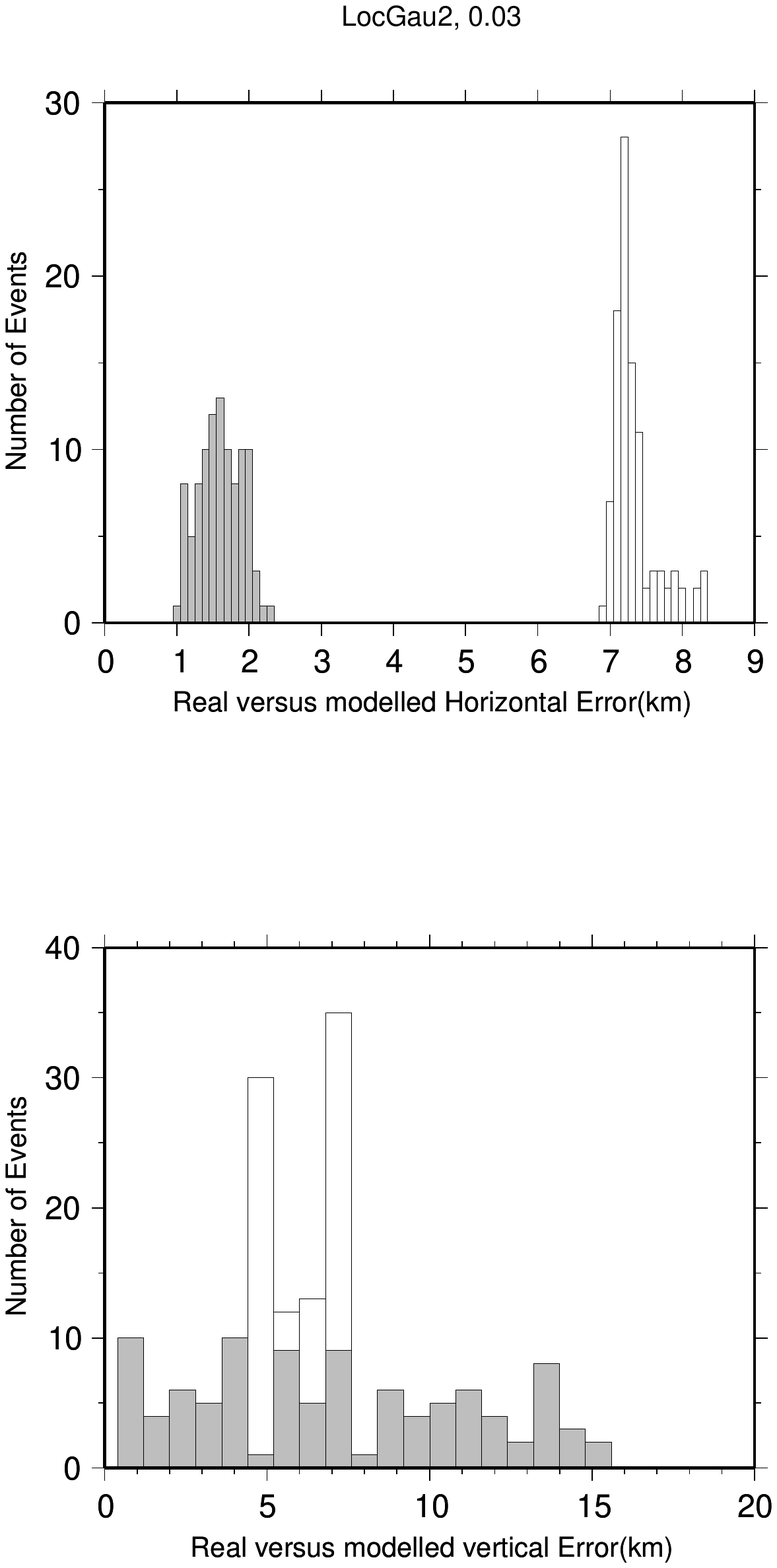}
\includegraphics[width=6.5cm]{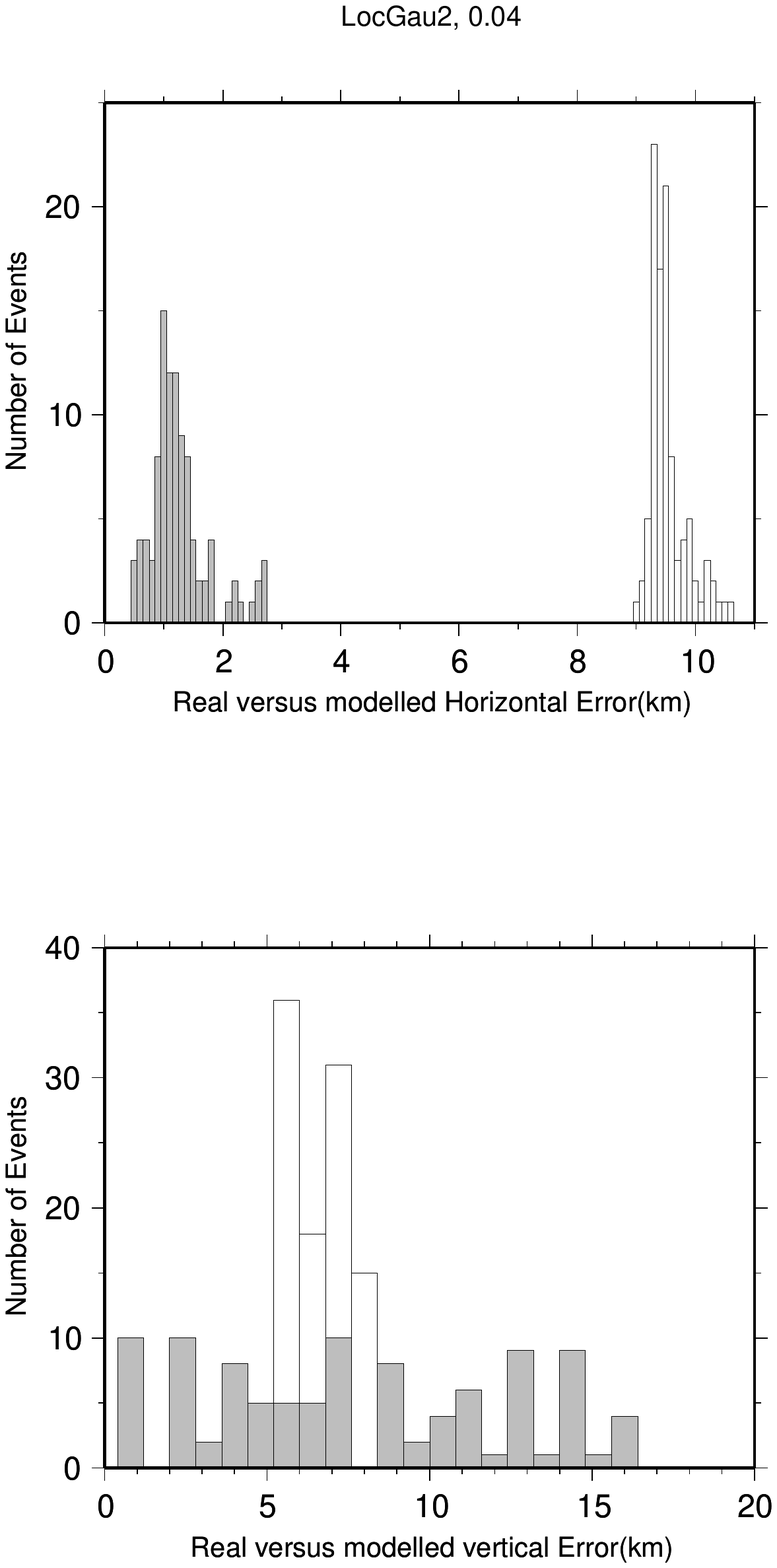}
\includegraphics[width=6.5cm]{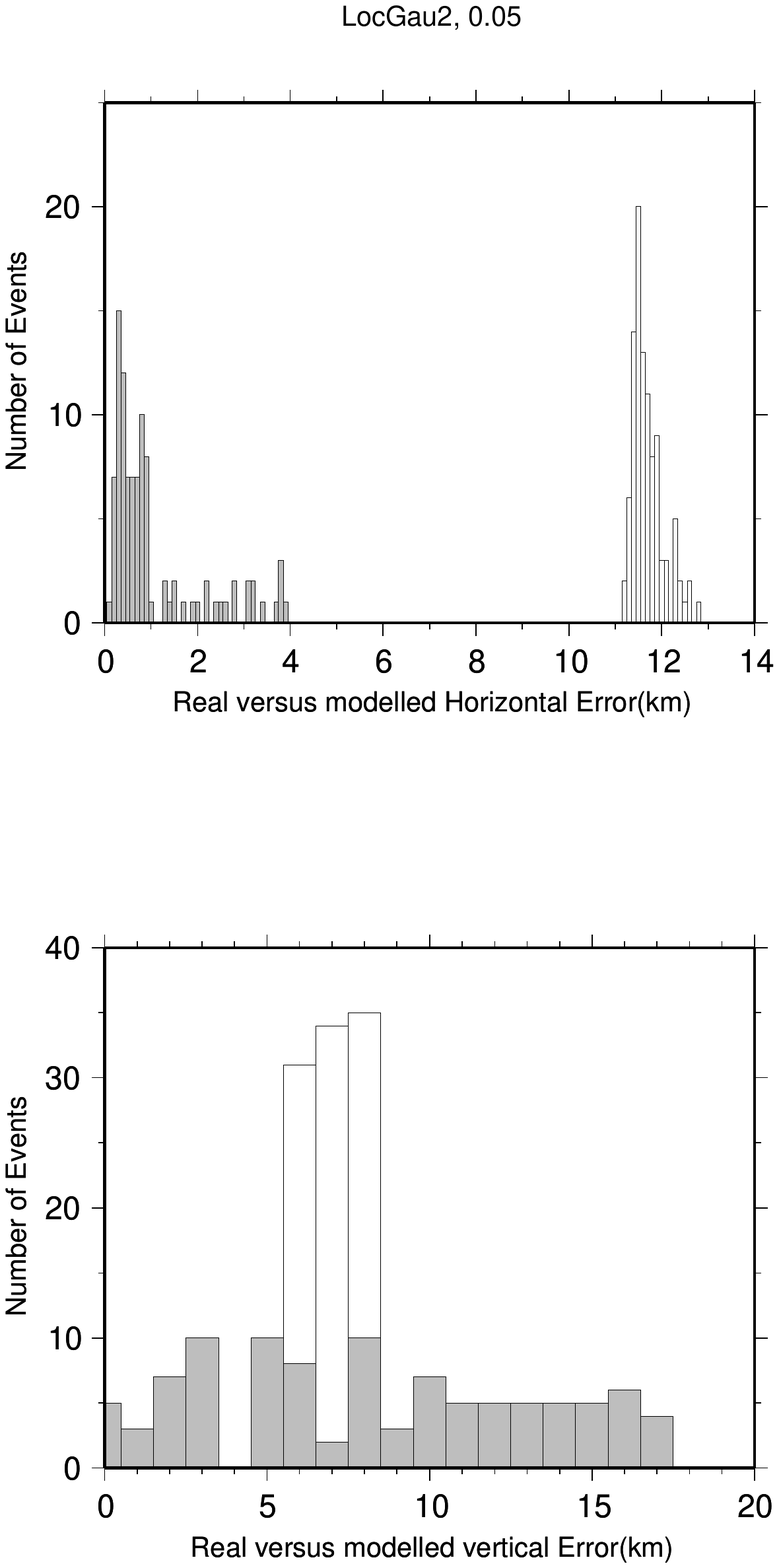}
\caption{\small {Histogram of real and modeled errors with different LocGau2 tries. Gray and uncolored bars show real and modeled errors respectively.}}
\end{sidewaysfigure}
Considering departure of modeled and real errors, LocGau2 = 0.01 seems to be a reasonable choice. Taking this value, relocation of all artificial events are performed, and the result is shown in the figure 5. It is worth to note that we have considered the average model as the final solution in the probabilistic distribution of all sampled models. Reason is the importance of the variation of probability of each model in the vicinity of the optimum one. To diminish unsystematic errors in the inversion process, considering that the variations of sampled model compared with optimum model is obligatory, since local maxima may mislead us for taking an untrue solution. This circumstance comes up when the final probabilistic distribution is not bell-shaped and would include some local maxima. Then, solution with maximum probability cannot certainly lead to the true solution, which is comparable with the real optimum model.

Noting the figure 5, some notes have to be put forward. First, we have succeeded to invert the epicenters with high accuracy and small real errors while trying to simulate real situation (with different forward and inversion earth models). Most of the events are relocated with the epicentral error smaller than 2 km. Another noticeable fact is that no events are relocated within the depth of 12 km (figure 5 E and F).

As seen in the figure 3, there is a high contrast between P-wave velocities in two models. IRSC velocity model features considerably higher P-wave velocities with respect to the Hatzfeld model (Hatzfeld et al. 2003). In consequence, sampled models tend to deeper locations to compensate for travel time difference that is present in the forward modeling using Hatzfeld velocity model. This fact shows the high impact of true velocity model of the earth to result in certain relocations.

Since the region of study is reliably similar to Hatzfeld velocity model, and the fact that the depth of sedimentary cover in the region is about 12 km (Shomali et al. 2011), we choose to exploit Hatzfeld model to relocate Shonbe seismic sequence. Although it would not be a reasonable choice for waveform modeling, as this velocity model is more like a local distribution of earth layers contemplating its geographical extent, Hatzfeld model comes with less details compared with IRSC one, then we will manipulate IRSC model for waveform modeling later in this study.
\begin{figure}[ht!]
\centering
\includegraphics[width=16cm]{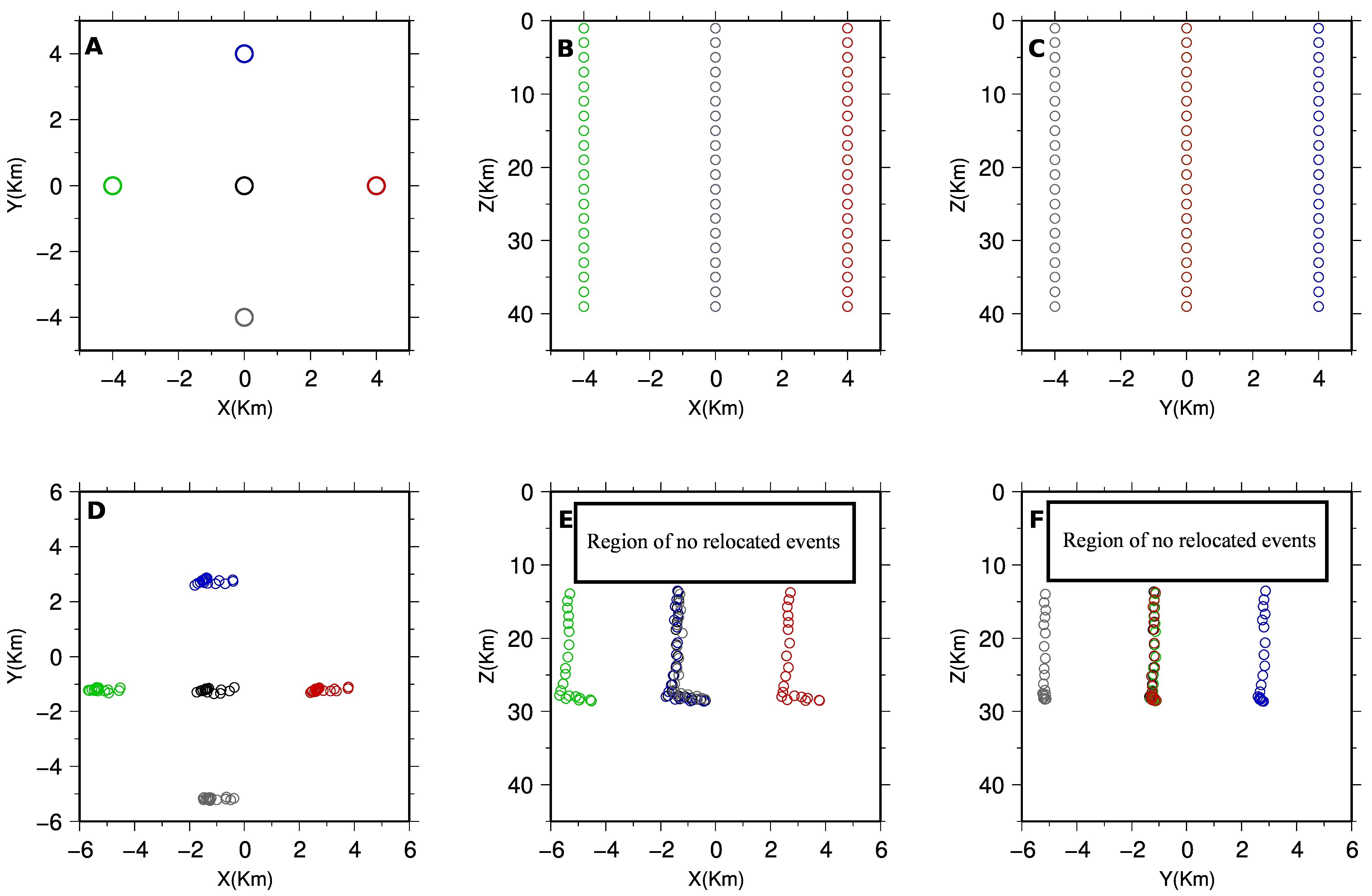}
\caption{\small {Relocation results of all artificial events with LocGau2 = 0.01. A, B, and C show events at their real positions versus D, E, and F showing relocated positions.}}
\end{figure}

\subsection{Relocation results}

 Looking at the figure 6 (A), most of the relocated events feature the epicentral error less than 5 km. Likewise, depth errors of most of the events are less than 5 km. Although the issue of true similarity between the Hatzfeld model and the real earth has remained unanswered. So, depth errors cannot be compared with real errors with high certainty. Considering the figure 6 (B), distribution of the stations can be investigated. Most of the relocated events feature azimuthal gap less than 160 to 180 degrees, because of the Arabian stations. Furthermore, no explicit deviation of secondary azimuthal gap compared with primary azimuthal gap is seen, suggesting uniform distribution of the stations and the fact that removal of any stations can considerably affect the results.
\begin{figure}[ht!]
\centering
\includegraphics[width=8cm]{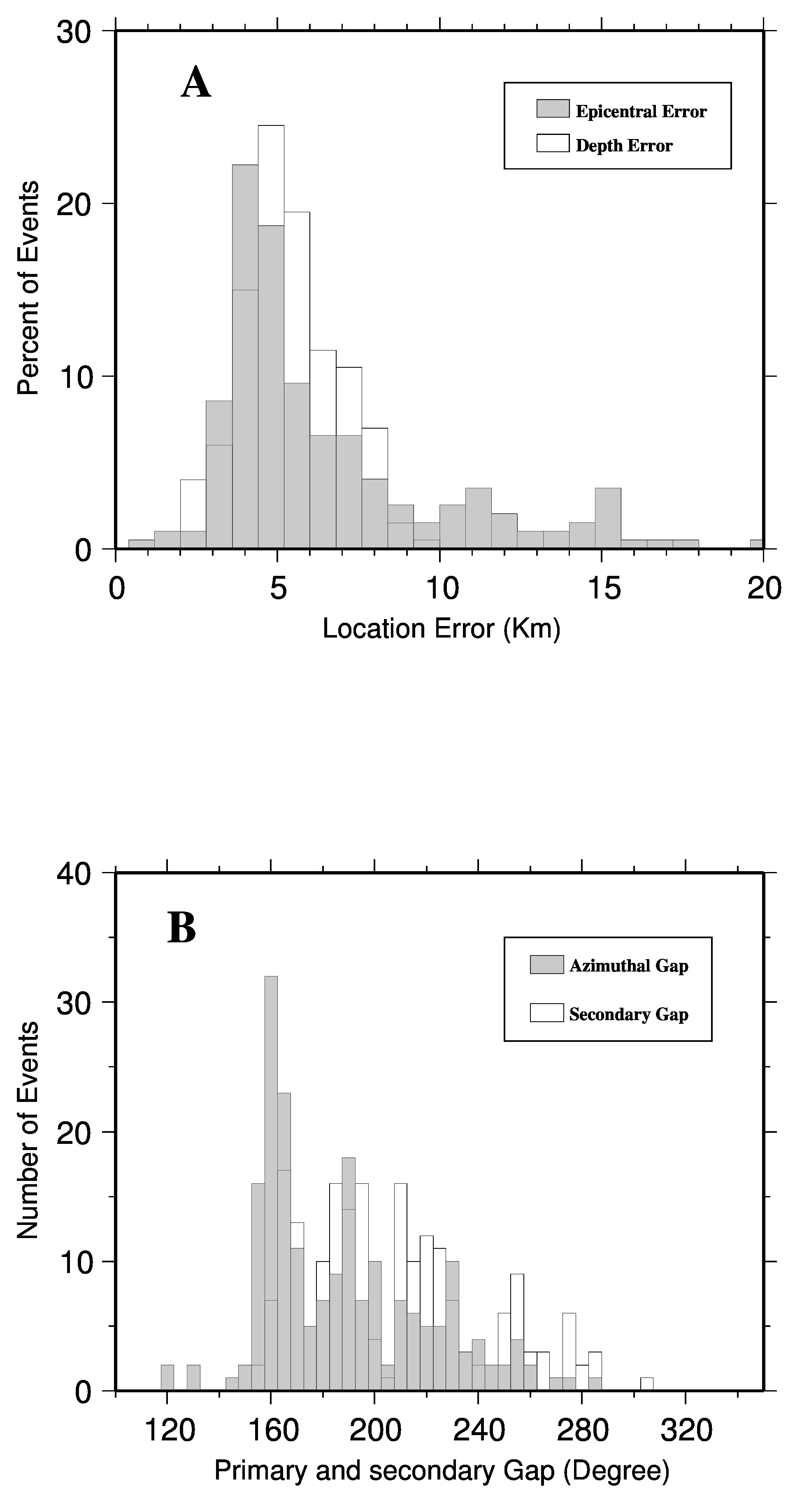}
\caption{\small {Histogram of epicentral and vertical errors accompanied by primary and secondary gaps for the relocated events. A shows the epicentral and vertical errors. B also depicts primary and secondary gaps that are relevant to the relocation results.}}
\end{figure}

In the figure 7, distribution of the sampled models for the main shock and its biggest aftershock is shown.  USGS and IRSC locations are depicted compared with our average model and most probable model in the figure. As explicitly seen, distribution of the models is mostly confined to 68-percent ellipse, which is relevant to the volume with 68$\%$ probability to happen as the hypocenter of an event. This alone shows that distribution of the sampled points based on their probability to happen is statistically Gaussian. In result, expressing the modeled errors using covariance matrix is reasonable and trustworthy. As mentioned before, average solution (instead of the most probable one) is the best choice as the output of a probabilistic distribution for each event, and one standard deviation can be regarded as the error of 68-percent probability.
\begin{figure}[ht!]
\centering
\includegraphics[width=13cm]{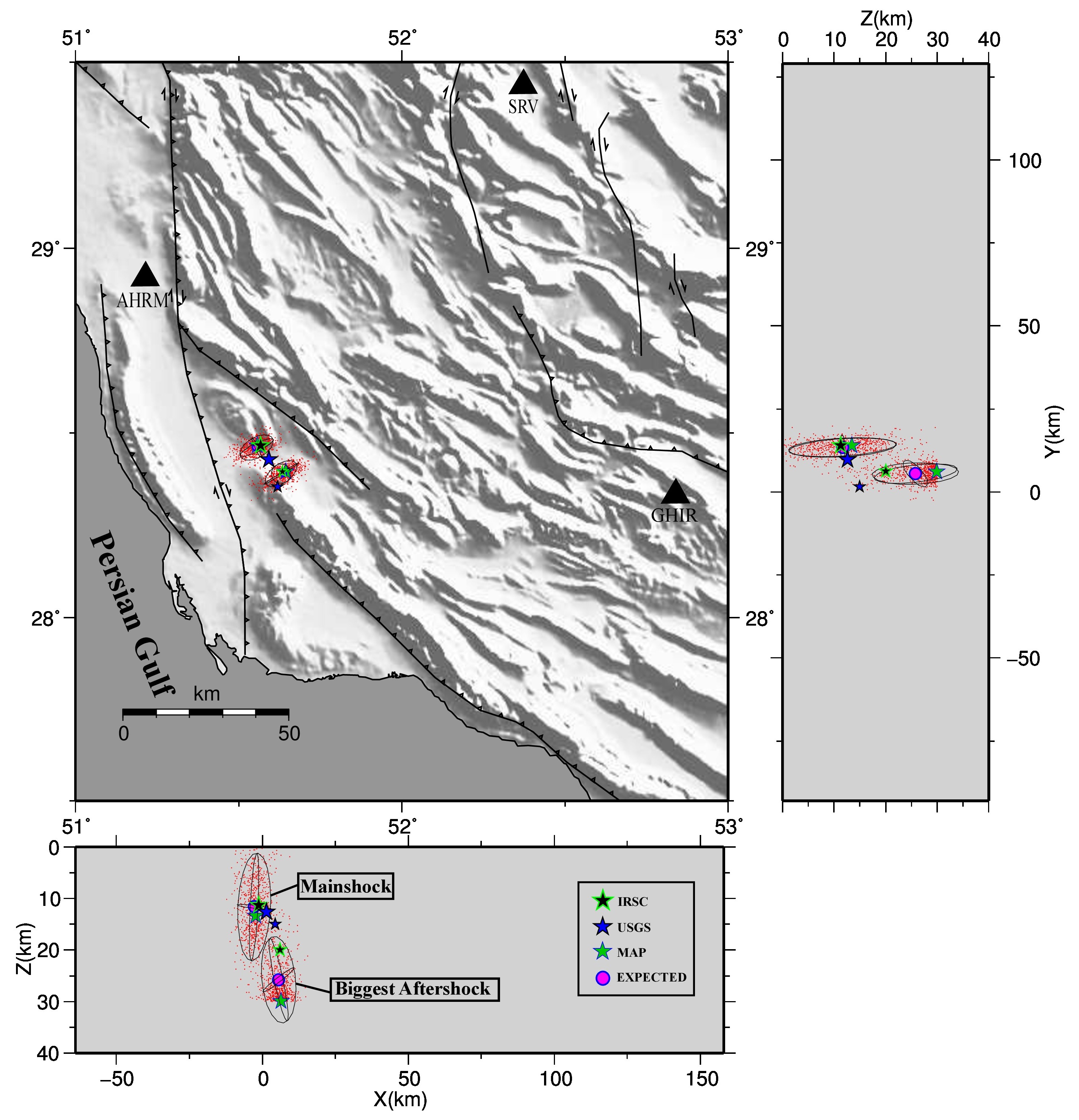}
\caption{\small {Distribution of the scattered points for the main shock and its biggest aftershock for acquiring the final probability density function.}}
\end{figure}

To grasp the temporal and geographical migration of the sequence, events with epicentral errors less than 5 km and with the time interval of up to 30 days after the main shock are colored based on their time of happening with respect to the main shock (figure 8). Main shock epicenter is relocated close to the Khaki anticline. Since this place is near Shonbe, damages may be considerable in this city as it was in reality. Considering spatial and temporal distribution of the aftershocks up to 30 days after the main shock, a northwest-southeast trend can be explicitly distinguished. Although much of the seismic activity is concentrated in the northwestern part, aftershocks move towards southeast by the time. But the southeastern part cannot be regarded as a new seismically activated region, because of the meaningful reduction of the aftershocks.

Noting the depth distribution of the aftershocks in the figure 8, concentration of events in the depth range of 15 to 30 km with steep dip (figure 9 (b)) can be seen, based on histogram presented. Considering many studies suggesting that the depth of sedimentary cover (Hessami et al. 2006) in ZSFB lies within 8 to 12 km (Tatar et al. 2004; Nissen et al. 2011; Yamini-Fard et al. 2006), occurrence of the aftershocks near the bedrock can be concluded. Also, main shock depth is 12 km implying that it occurred in the upper-most part of the bedrock in the vicinity of sedimentary cover. To verify this claim, it is worth to note that the depth range of seismic events in ZSFB is almost concentrated in the bedrock (Jackson 1980; Jackson and Fitch 1981). Furthermore, rupture would be perhaps aseismic in the sedimentary cover, for instance shortening, hence occurrence of the aftershocks in the bedrock is certainly reasonable (Kadinsky-Cade and Barazangi 1982; Maggi et al. 2000).
\begin{figure}[ht!]
\centering
\includegraphics[width=13cm]{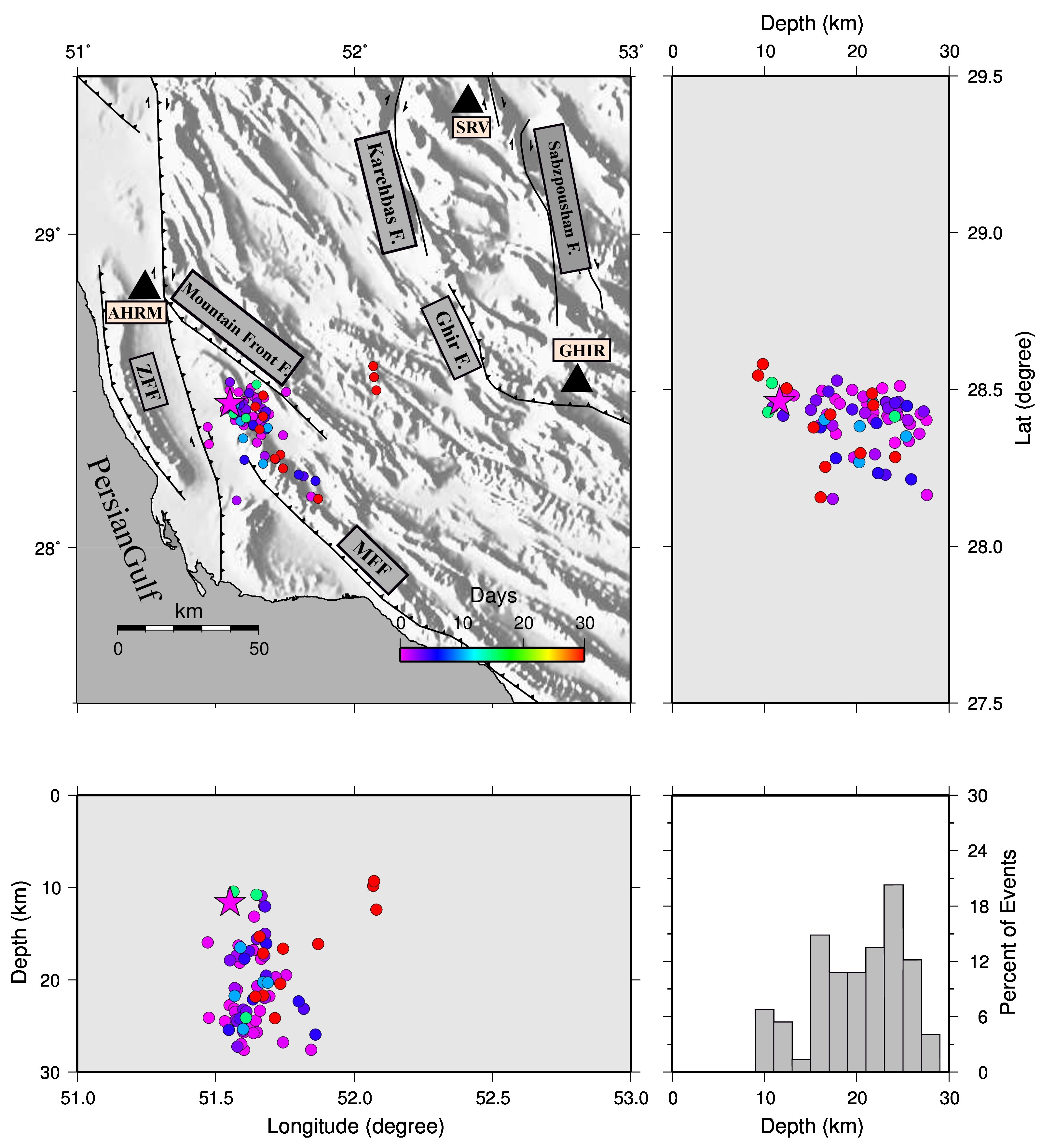}
\caption{\small {Colored events within 30 days after the main shock accompanied with the histogram of their depths.}}
\end{figure}

To investigate spatial migration of the events further, two cross sections with azimuths of 60 and 135 degrees are taken in the figure 9 (a) and (b). All the relocated events are projected onto these cross sections, in result a steep dip of about 70 degrees for the probable causative fault is seen (figure 9 (a)). This dip is in accord with 50-60 degrees of most of the seismically active faults in ZSFB. To be added, migration of the aftershocks to southeast and upper-most part of the bedrock can be concluded (9 (b)).
\begin{figure}[ht!]
\centering
\includegraphics[width=13cm]{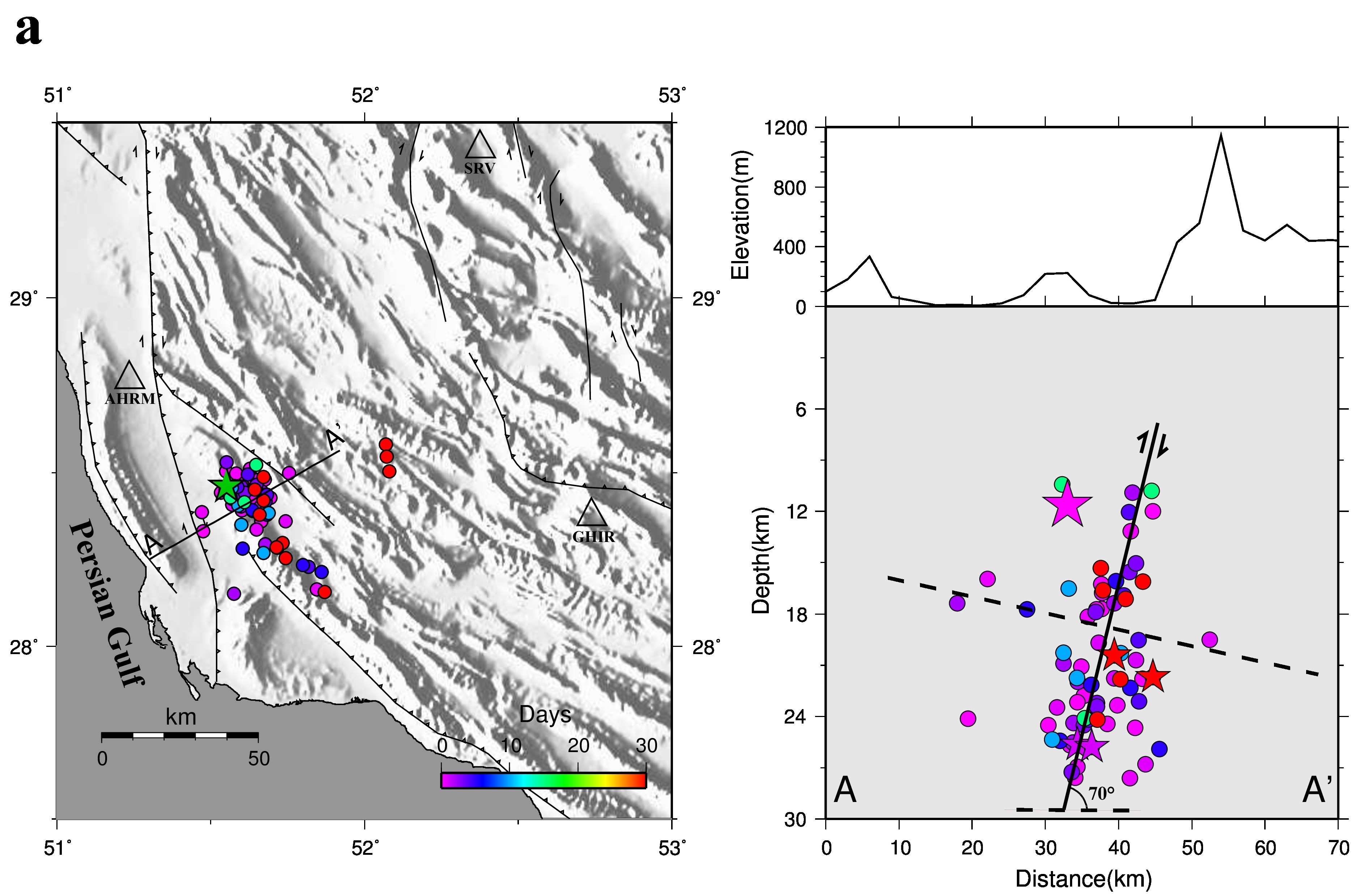}
\includegraphics[width=13cm]{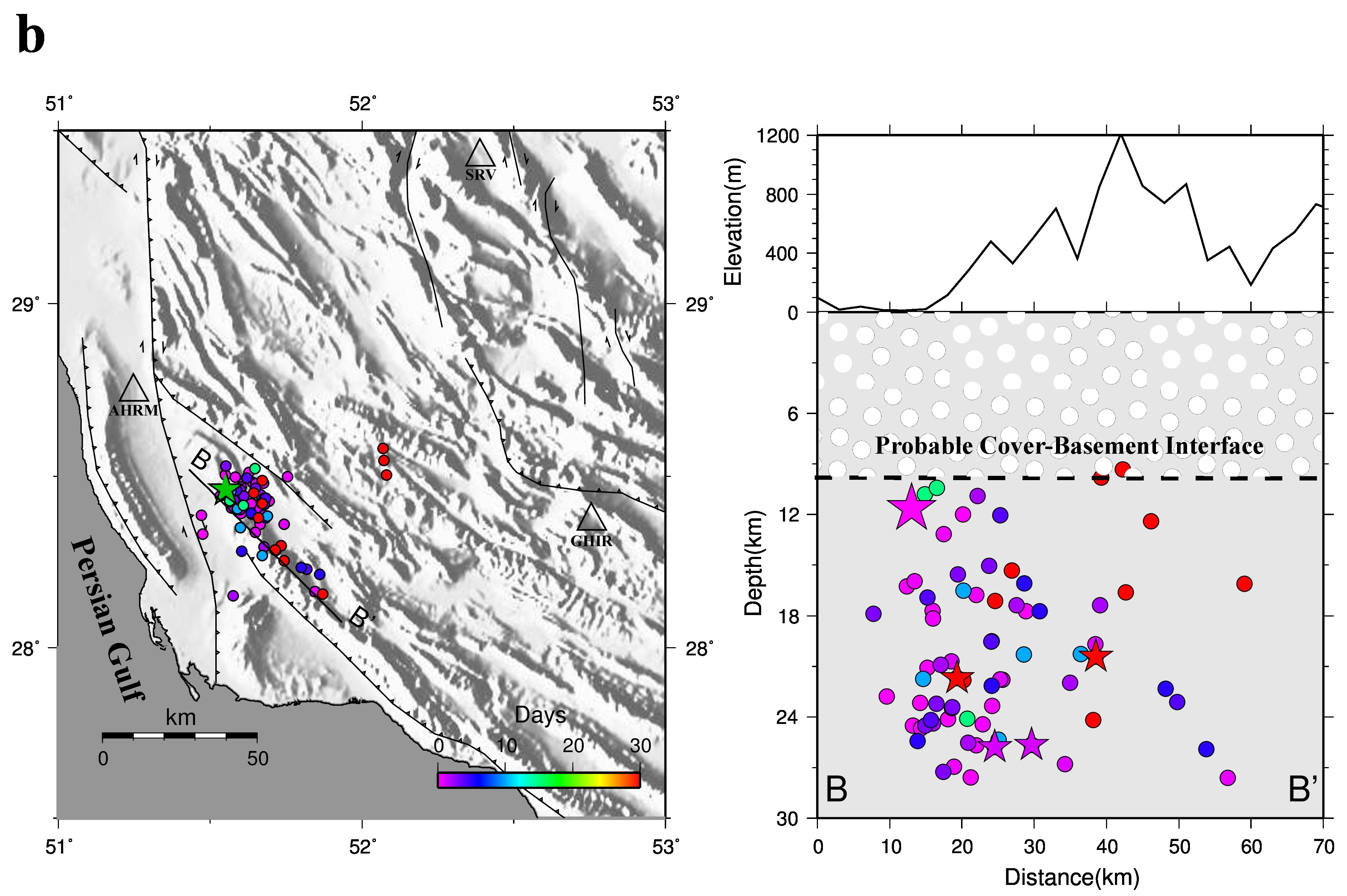}
\caption{\small {Relocated events within 30 days after the main shock onto two cross sections of (a) 60 and (b) 135 degrees. The big purple star represents the main shock, and the other stars are showing the biggest aftershocks.}}
\end{figure}

\section{Source waveform modeling for some big events}

Focal mechanism inversion for 21 seismic events with magnitude greater than 4.5 Mn is performed with Zaheradnik method (Zahradnik et al. 2005; Zahradnik, Gallovic et al. 2008; Sokos and Zahradnik 2008; Zahradnik, Jansky et al. 2008; Zahradnik and Custodio 2012; Sokos and Zahradnik 2013) and regional waveform data. Considering the advantages and output possibilities, we exploit the new version of ISOLA (Zahradnik, Gallovic et al. 2008) to acquire the results. Furthermore, sub-event investigation for the main shock is performed, and stability of the acquired focal mechanisms is probed for 10 chosen events from the all 21 inverted events.

\subsection{Data and velocity model}

Waveform data of wide-band stations of IGUT and IIEES are used for the inversion task. Almost different stations are considered for inversion of each event noting data quality and SN ratio. Data of AHRM (Ahram), JHRM (Jahrom), SHI (Shiraz), and GHIR (Ghir) are mostly used for the events. Furthermore, Wavenumber method of Bouchon (Bouchon 1981) is utilized to calculate Green’s functions. IRSC velocity model is employed for the calculation of Green’s function noting that this model leads to less travel time residual times at distant stations and the most variance reduction of the real waveforms compared with more local Hatzfeld model. To be added, inversion is performed in two steps. First, we seek the best position in a vertical grid with the epicenter relocated formerly starting at 1 km depth with 1 km vertical interval between each node for the main shock and 2 km vertical interval for other inverted shocks (last node is at 30 km depth). Secondly, we continue to chase within a horizontal grid at chosen depth in the previous step with 25 nodes that are placed differently for each event. All the waveforms of the stations have band-pass filtered in a different frequency range, to reach more data quality and variance reduction.

\subsection{Inversion results}

In this section, all the acquired results of the point-source inversion of the main shock are presented. Error estimation and also stability of the results are examined. In advance, we try to present our interpretations about the rupture geometry of the sequence. Finally, all the results with their error estimation are presented for 10 chosen events to reach an interpretation about the rupture process within the region of impact. Main shock of this seismic sequence is relocated at 51.55 East Longitude, 28.46 North Latitude at 11:52:48 UTC in the border of Khaki anticline close to Shonbe city, based on the results of nonlinear method.  Considering the moment magnitude of the main shock (6.4 Mw), it may seem reasonable to have complexity in the rupture process, then there may exist a sub-event and then inconsistent point-source assumption.
\begin{figure}[ht!]
\centering
\includegraphics[width=13cm]{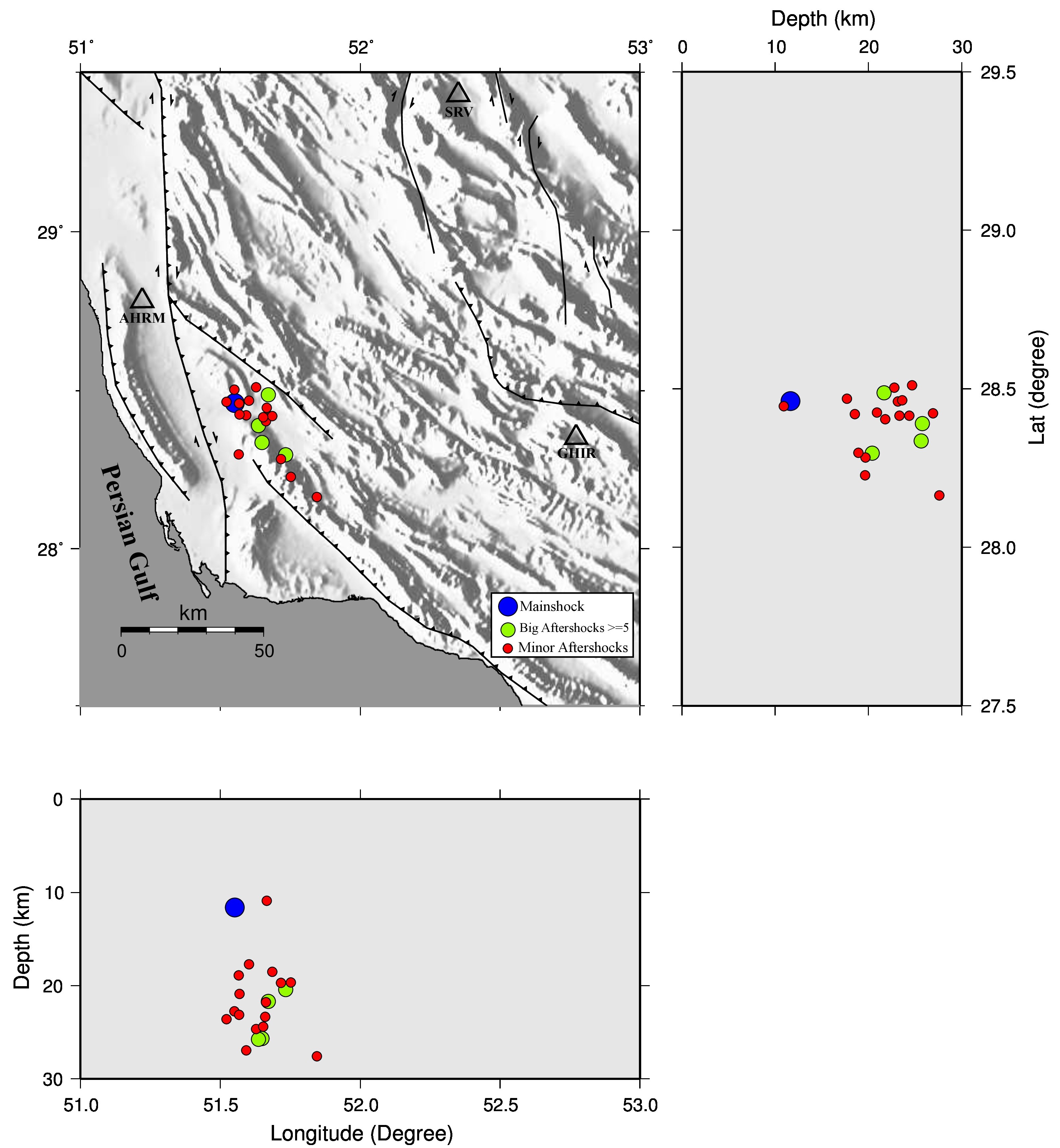}
\caption{\small {Chosen events for the point-source inversion with Mw\,$\geq$\,4.5.	}}
\end{figure}

First, seismic source inversion is performed assuming a single event for the main shock, in advance this assumption will be investigated carefully for reaching more certainty of the results and the resulting analysis. Considering different qualities of recorded data and existing noise for each station, chosen stations are shown in the figure 11.
\begin{figure}[ht!]
\centering
\includegraphics[width=13cm]{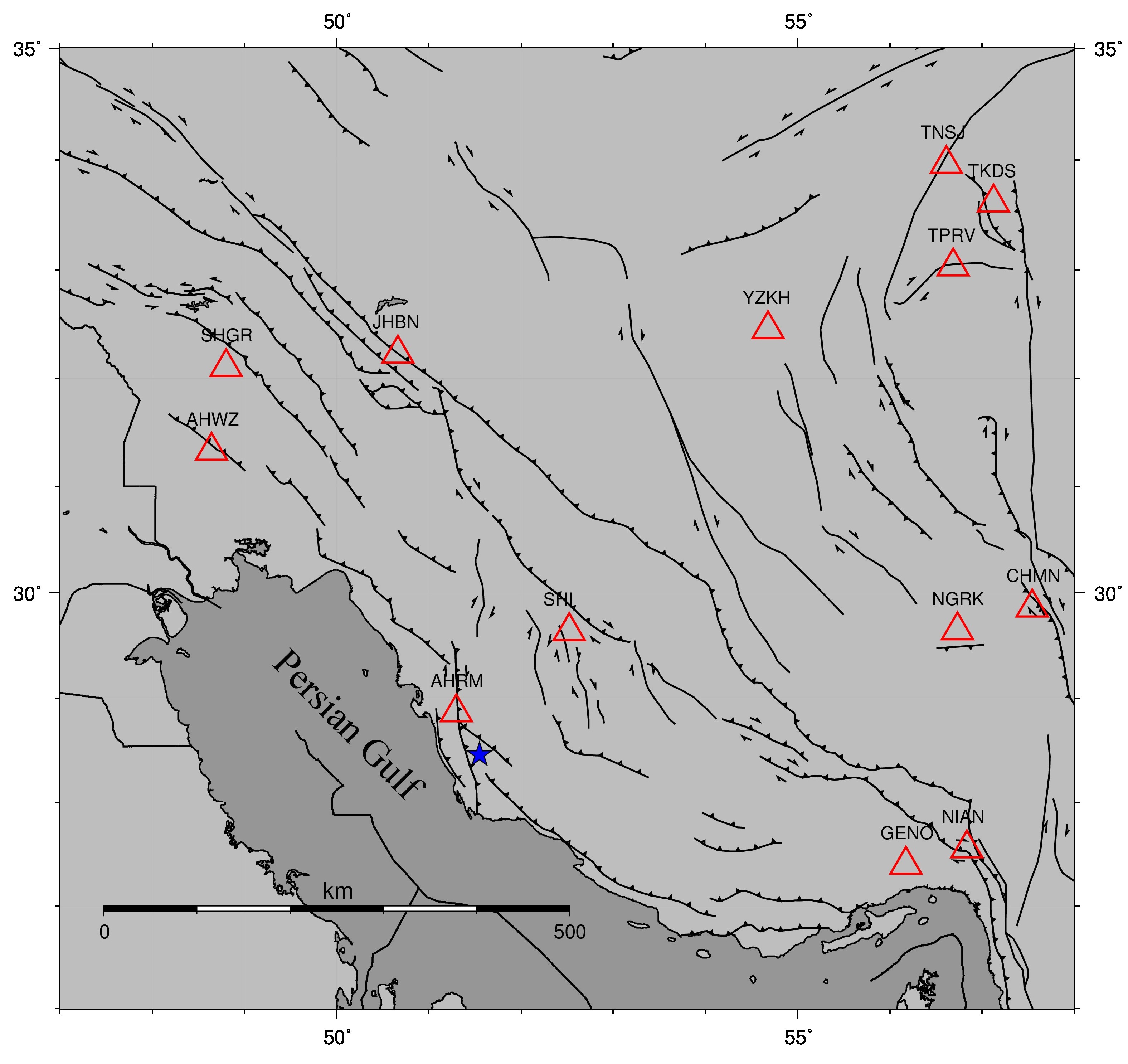}
\caption{\small {All the stations used for main shock forward waveform modeling.	}}
\end{figure}

Comparing calculated Green’s functions with band-pass filtered recorded data in different frequency ranges, appropriate frequency range for each station is retrieved and shown in the table 1.
\begin{figure}[htb!]
\centering
\includegraphics[width=8cm]{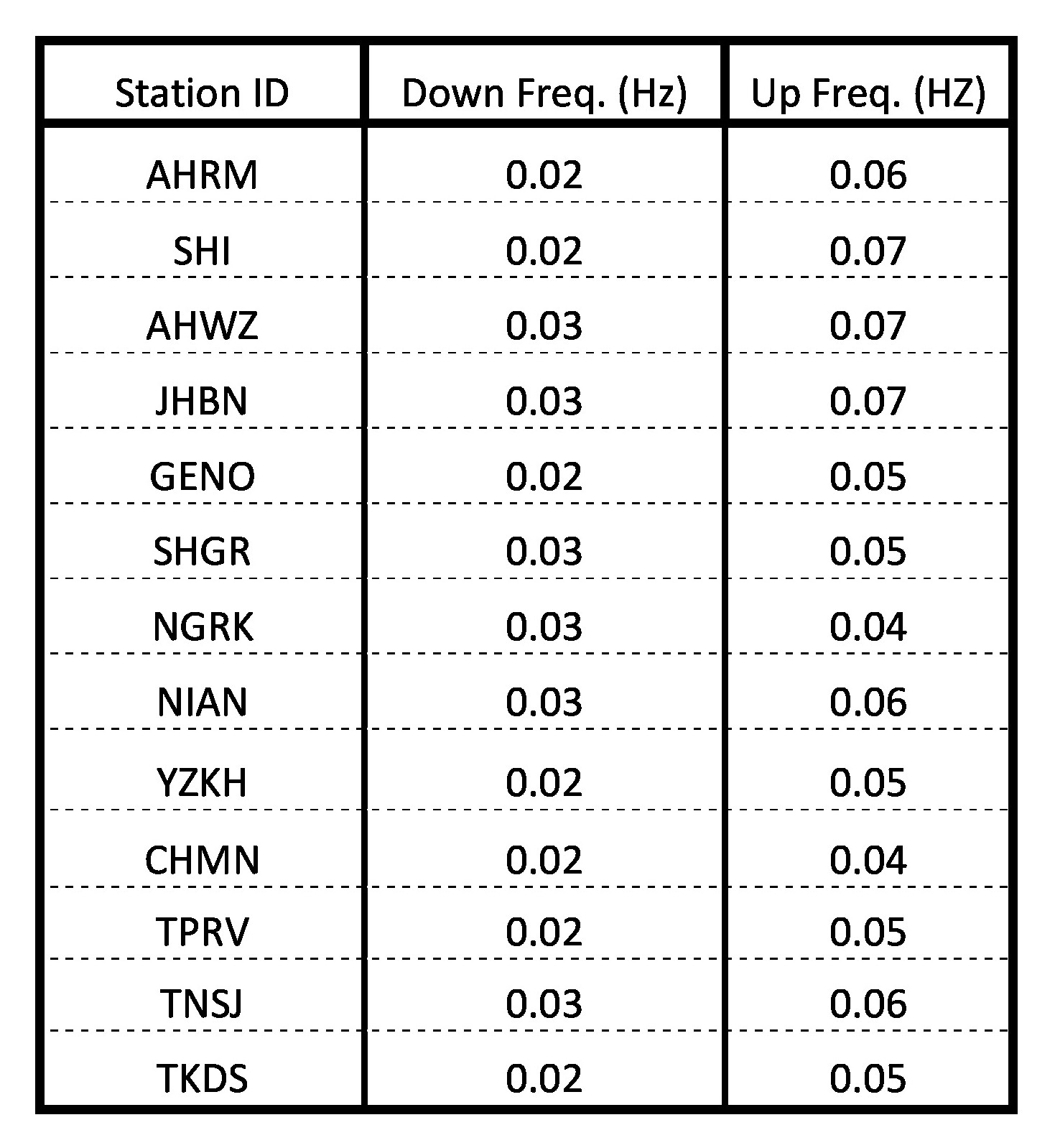}
\caption*{\small {Table 1: Frequency range suitable for each station considering the result of waveform band-pass filtering.}}
\end{figure}

To commence the inversion, one vertical grid at the relocated epicenter with 15 nodes placed in 1 km depth intervals started at 5 km and ending at 19km is established. After finding the best node, a horizontal grid of 25 nodes with 2 km intervals has been regarded.
\begin{figure}[htb!]
\centering
\includegraphics[width=11cm]{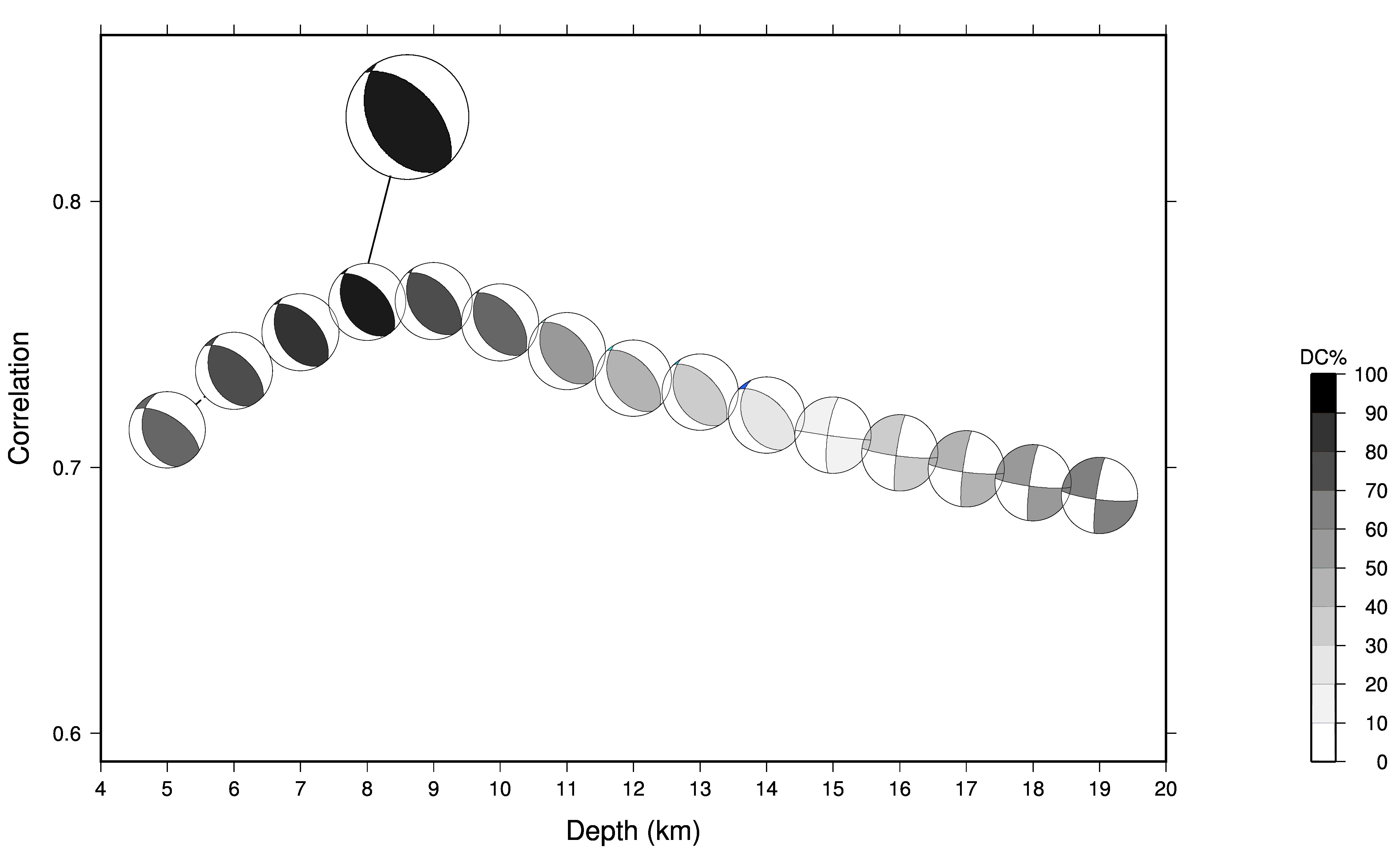}
\caption*{\small {Figure 12: Inversion result of the main shock focal mechanism within the vertical profile.}}
\end{figure}

\begin{figure}[htb!]
\centering
\includegraphics[width=10cm]{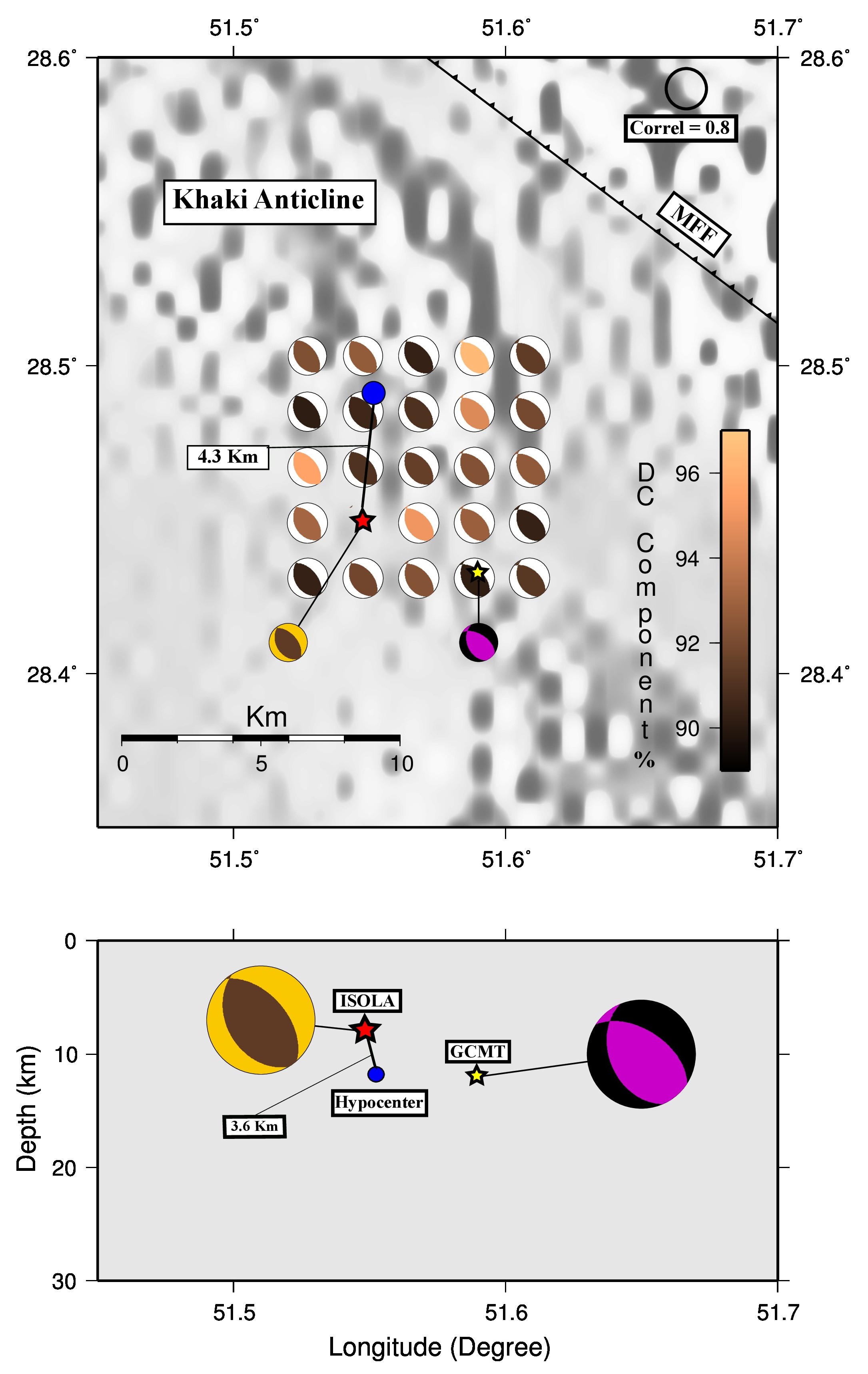}
\caption*{\small {Figure 13: Inversion result for the main shock focal mechanism within the horizontal grid placed at 8km depth. Focal spheres are colored based on DC value and their size that suggests the correlation between observed and modeled waveforms. GCMT results are also shown for comparison.}}
\end{figure}
Regarding the figure 13, horizontal distance between inverted centroid and the relocated epicenter is 4.3 km. To be added, inverted source geometry presents a reverse slip vector with a minor strike-slip component. Furthermore, inverted depth of centroid is 8 km that presents the initiation of the sequence at the lower-most part of the sedimentary cover. Considering the horizontal position of centroid compared with the relocated hypocenter, spread of the sequence to the southeast can be viewed explicitly in accordance with former relocation and GCMT results.
\begin{figure}[htb!]
\centering
\includegraphics[width=16cm]{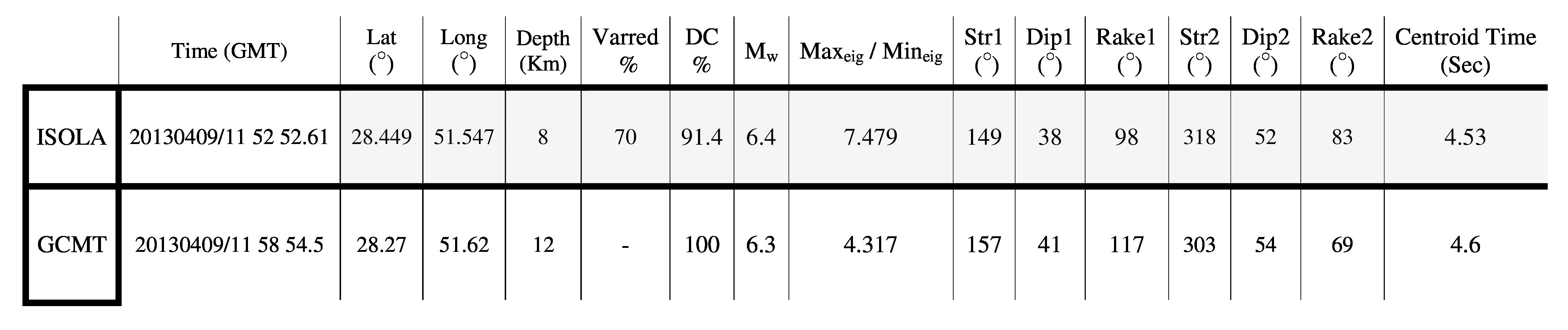}
\caption*{\small {Table 2: Point-source inversion result for the main shock in comparison with GCMT result.}}
\end{figure}

\subsection{Sub-event inspection}

Being aware of the causative fault geometry, we intend to investigate any possible complexity in the rupture process. Hence, an oblique fault plane grid with $151^{\circ}$ strike and $36^{\circ}$ dip including 25 nodes that are placed with 0.5 km intervals in dip direction and 2 km intervals in the strike direction is established. This grid can somehow simulate the true fault plane. The origin node of the grid is the centroid that has been acquired before, and is placed as the first and second nodes in the dip and strike directions orderly. In this grid, all the nodes will be inverted for the possible sub-event of the main shock.

In the table 3, inversion task is performed in three steps across the possible causative fault plane. First, table 3 (a), inversion is executed assuming single event and deviated moment tensor that lead to an acceptable variance reduction and considerable double-couple component and seismic moment magnitude. Regarding these three factors, any possible sub-event can be inspected carefully.

Secondly, table 3 (b), one possible sub-event is assumed in the inversion. Nevertheless, variance reduction alters scarcely, and inverted slip vector is not compatible with the main event (normal component compared with the reverse slip of the main event). In consequence, the assumption of having second sub-event for the shock is rejected. Also this inspection is performed for two sub-event possibility (table 3 (c)), that again is declined because of the same mentioned argument. After this examination, we intend to investigate the stability and uncertainty of the solution assuming single-event main shock.
\begin{figure}[htb!]
\centering
\includegraphics[width=16cm]{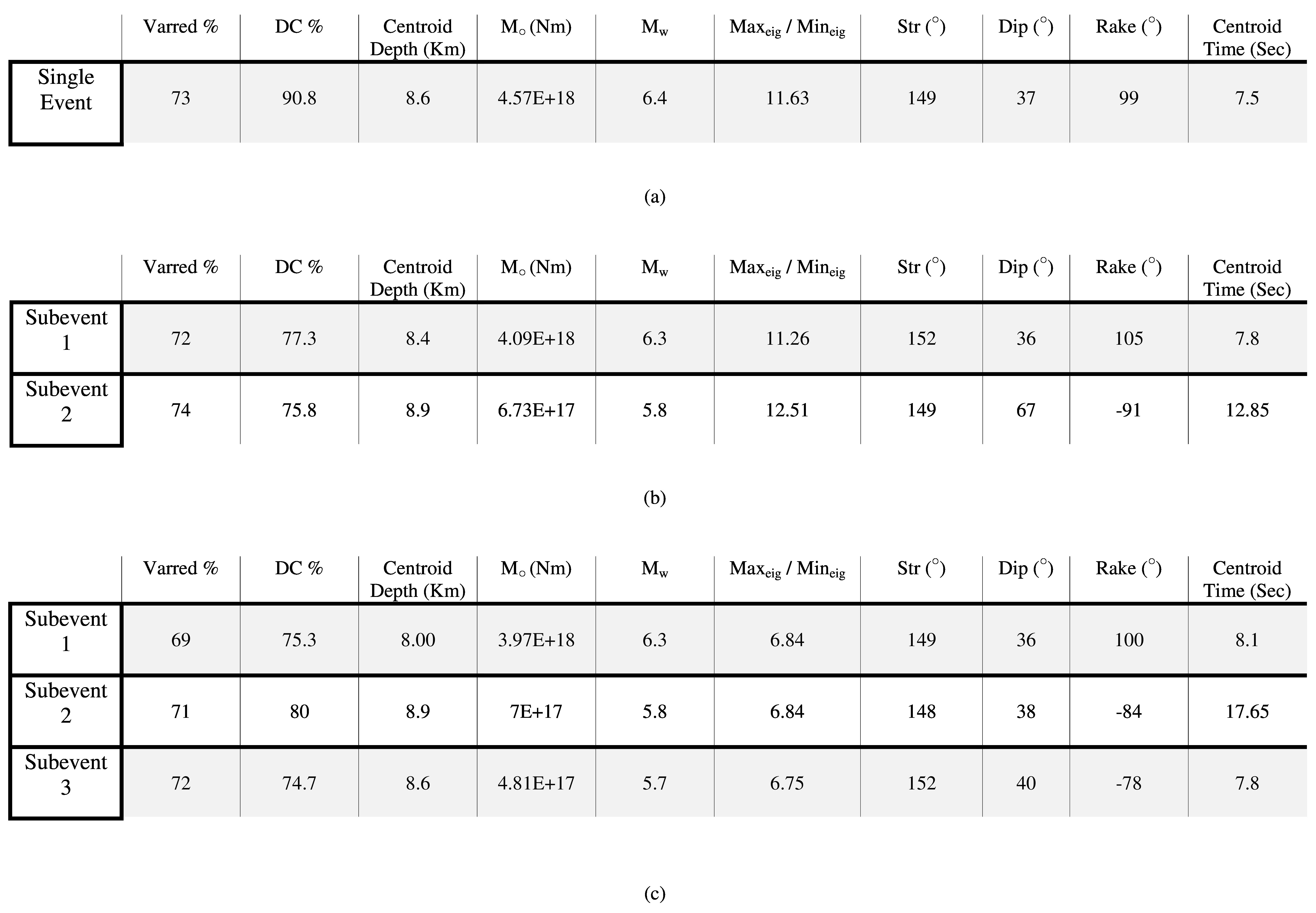}
\caption*{\small {Table 3: Inversion results of the main shock assuming one, two, and three sub-events considering an oblique grid with strike and dip similar to the probable causative fault. Results suggest that the main shock comprises single event.}}
\end{figure}

\subsection{Stability and uncertainty of the acquired model}

Considering different qualities of recorded data and existing dissimilarities between the one-dimensional velocity model and the true complex velocity structure, seismic source inversion can be highly affected by removing one or more stations from the inversion process. We expect that the removal of the stations with the most disparity of the velocity structure compared with one used in the inversion can certainly impact the results. Now, we intend to remove each station, one by one, and investigate the consequent results. It is expected to acquire the most affected result when the stations with either the highest or the lowest correlations are removed, because of the use of $L_2$ norm in the correlation calculation. Results are then shown as histograms in the figure 15.
\begin{figure}
\centering
\includegraphics[width=10cm]{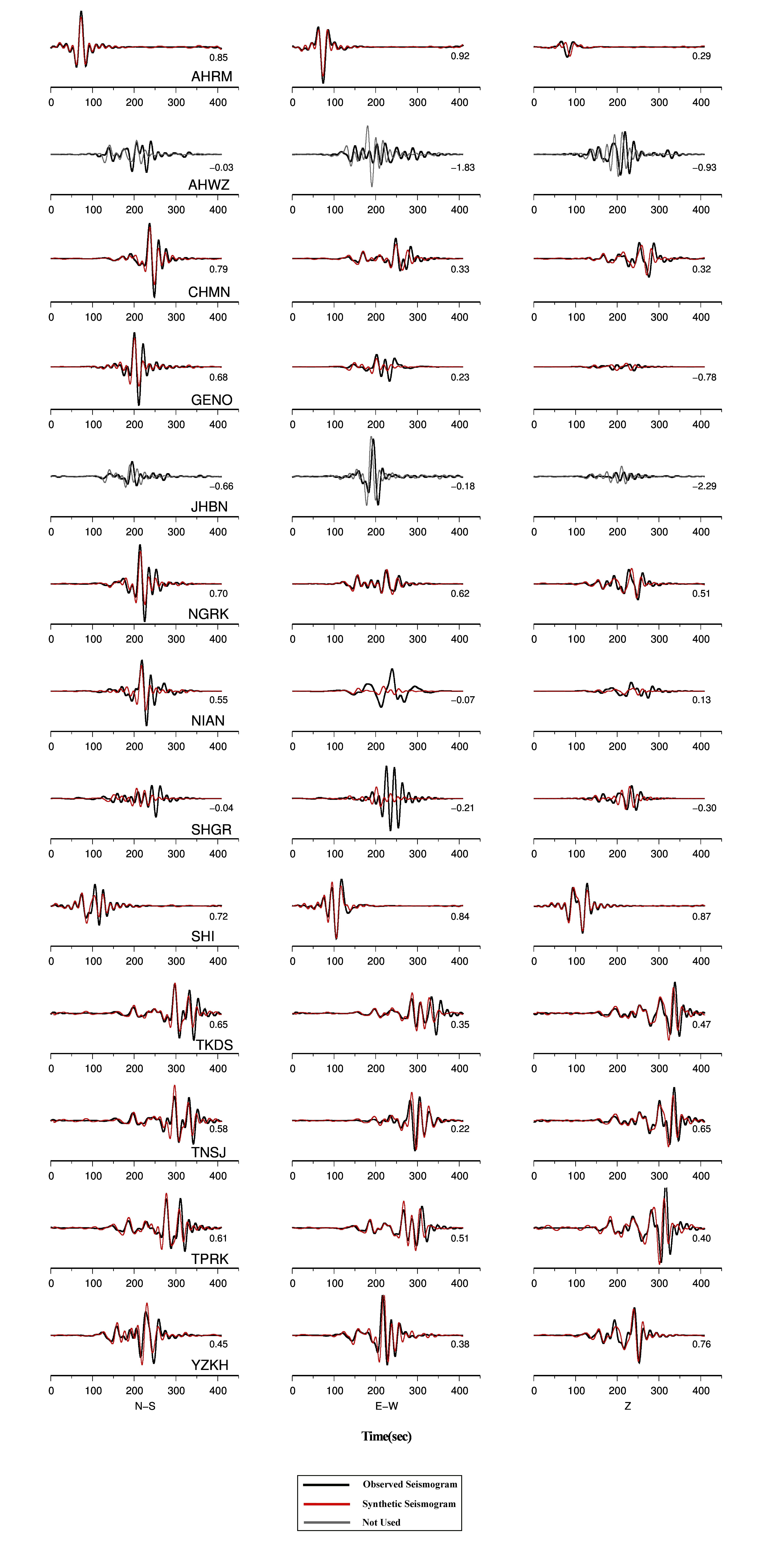}
\caption*{\small {Figure 14: Observed and modeled waveforms resulting from the main shock point-source inversion. The numbers inside the boxes are correlations between calculated and observed waveforms.}}
\end{figure}

Regarding the figure 15 (e), slip vector shows a reverse motion since all the inverted slip vectors for two possible fault planes present such this motion. In figure 15 (f), suspected some complexity of the rupture process is noticeable regarding little DC components retrieved in the inversion for each step.
\begin{figure}[htb!]
\centering
\includegraphics[width=16cm]{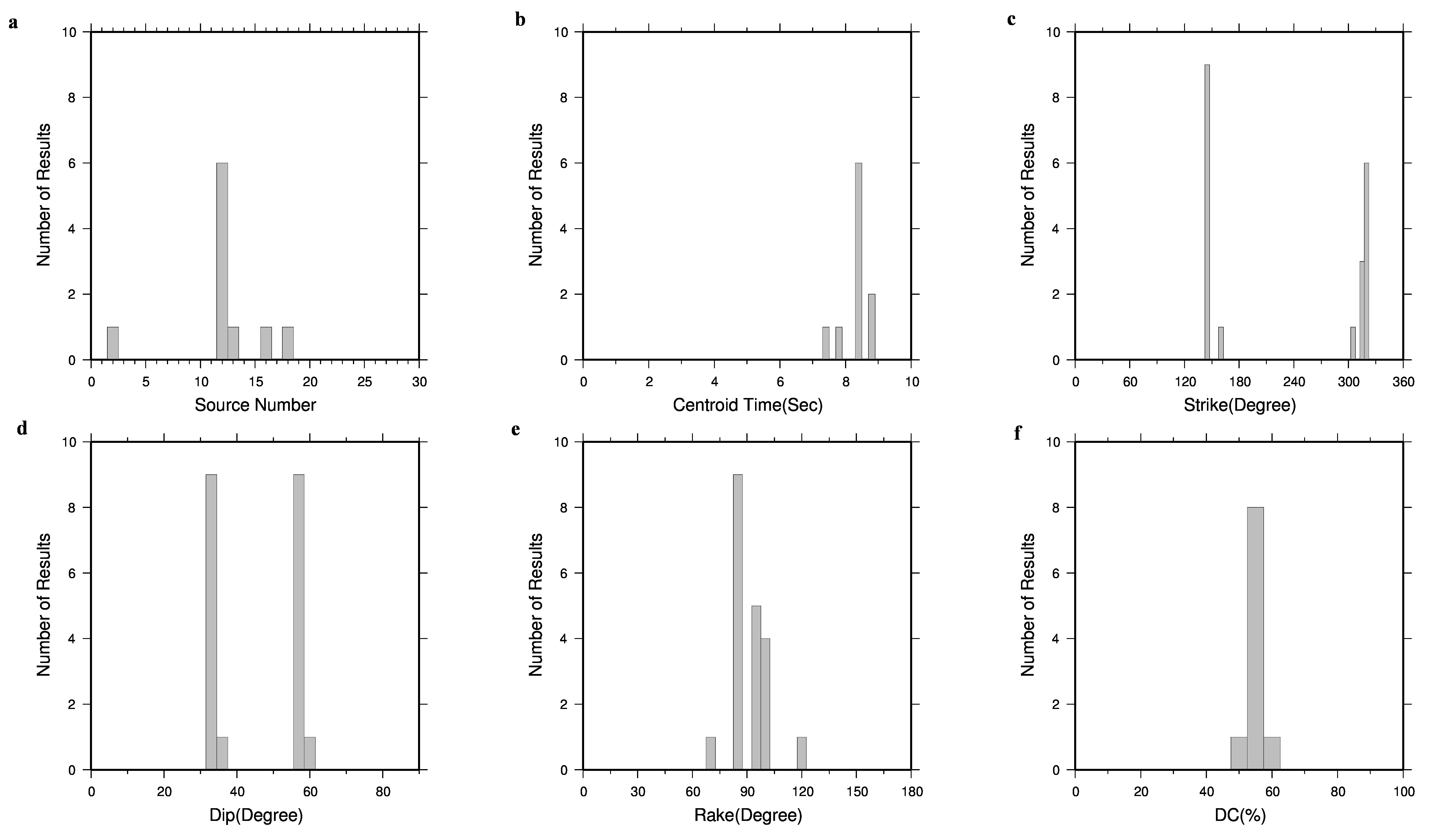}
\caption*{\small {Figure 15: Results of the stability test for final inversion results. One station is removed in each step.}}
\end{figure}

Nevertheless, more inspections show very little difference in the value of variance reduction and correlation for the results with much more DC components. Also removing AHRM station, which has the most correlation with the calculated waveforms, lowers DC component considerably, results in the low DC that is nonsense.
\begin{figure}[htb!]
\centering
\includegraphics[width=12cm]{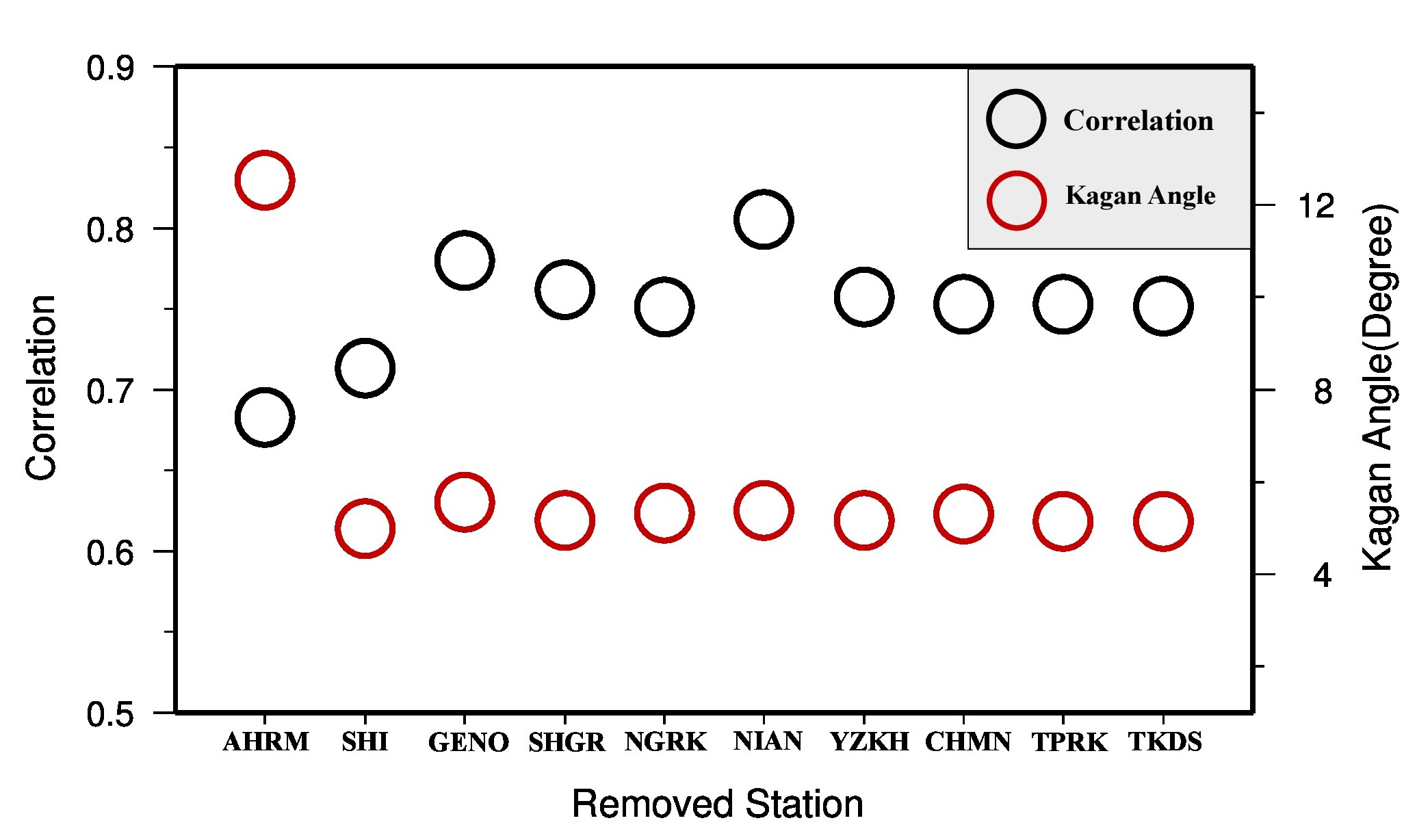}
\caption*{\small {Figure 16: Distribution of all correlation values for each step with the relevant Kagan angle.}}
\end{figure}

Regarding the figure 16, fluctuation of the correlation between observed and calculated waveforms is not considerable except for two stations AHRM and SHI. As the substantial impact of these two stations upon the final variance reduction, removing each one from inversion process can highly affect the results. As mentioned before, the function to calculate variance reduction is $L_2$ norm. It means that removing the stations with the low correlation can lead to considerable increase of final variance reduction, as is seen for the stations NIAN, SHGR, and GENO.

Moreover, low Kagan angle (Zahradnik and Custodio 2012) variances can be viewed in the figure 16, hence it shows the little changes of the inverted source parameters and consequent stable optimum parameters. Nevertheless, removing AHRM data leads to $12^{\circ}$ Kagan angle. This again presents the substantial impact of removal of a station with the most effect on the final variance reduction and the resulting source parameters. Considering high correlation between calculated and observed waveforms for this station (figure 14), its removal certainly leads to the considerable change in the inverted source parameters and a suitable variance reduction. In the figure 17, all the inverted source parameters are shown on the focal sphere for the  removal of each station separately.
\begin{figure}[htb!]
\centering
\includegraphics[width=4cm]{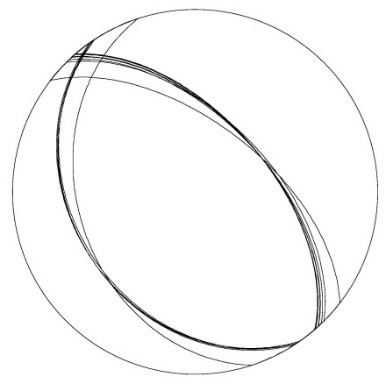}
\caption*{\small {Figure 17: Point-source solution on the focal sphere for each step of stations removal.}}
\end{figure}

\subsection{Error estimation of the source parameters without waveform inversion}

In this section, we attempt to determine stability of the acquired source parameters employing Zahradnik and Custodio method (Zahradnik and Custodio 2012), which is to reach the estimation without any further inversion of the models in vicinity of the results. This method seems rapid and efficient because of the inspection of too many models in vicinity of the optimum results. First, a threshold value for variance reduction must be set. Afterwards, all the double-couple moment tensors are retrieved as resulting variance reduction in the range of threshold value and optimum value. To better grasp the distribution of these moment tensors in the model space, converting to geometrical source parameters is executed.
\begin{figure}[htb!]
\centering
\includegraphics[width=16cm]{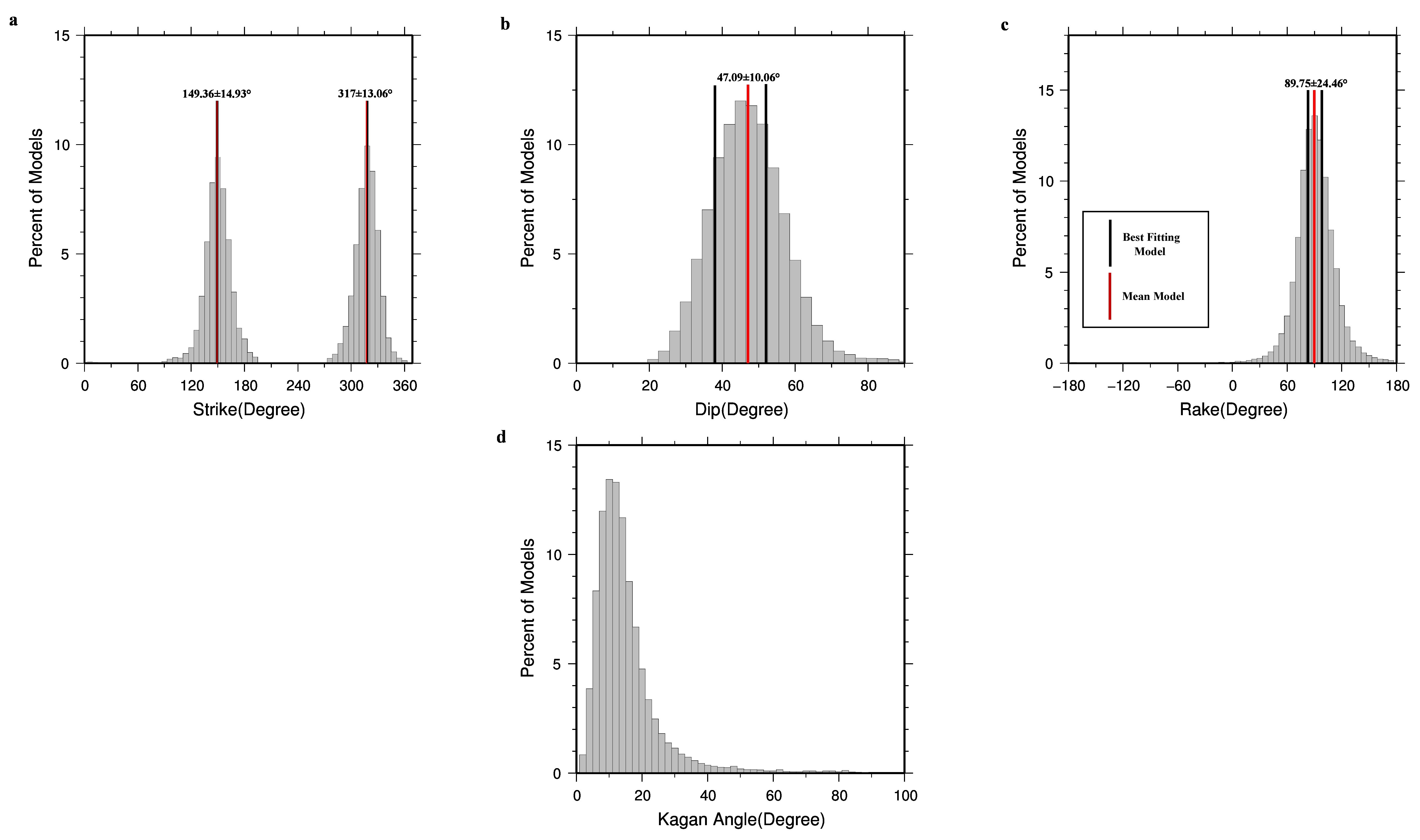}
\caption*{\small {Figure 18: Histogram of the models parameter in the vicinity of the optimum solution. 10 percent difference of correlation is considered to specify the sampled models. Also Kagan angle histogram of the picked models is shown (d).}}
\end{figure}

C is set 2 (relation (113) in Zahradnik and Custodio (2012)), and we note that the real data error threshold value of variance reduction is acquired by setting $\Delta\,X^2 = 1$ (relation 15). Afterwards, all the DC models in the retrieved variance reduction interval are obtained and converted to the relevant geometrical source parameters. All the resulting models are shown on a focal sphere and are depicted also as the histograms. It is interesting to note that less scattering of the models in vicinity of the optimum model is expected, then larger value of variance reduction interval, in the case of stable inverted optimum model. As is seen in the figure 18 (a-d), distribution of the source parameters relevant to the models in vicinity of the optimum models presents a Gaussian behavior that implies the exploitation of standard deviation to estimate error of the resulting models. For dip and slip vectors, one pick can be distinguished since two fault plane solutions are close. In result, mean value solution deviates from the optimum models of two fault planes. Furthermore, mean value of the Kagan angle is also calculated as 14 degrees. It is important to note that distribution of the models around the optimum model is acquired by use of 10 percent of variance reduction interval compared with the best solution.
\begin{figure}[htb!]
\centering
\includegraphics[width=4cm]{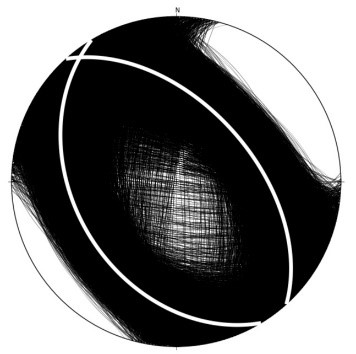}
\caption*{\small {Figure 19: Focal mechanisms of all 69712 DC models in vicinity of the optimum model.}}
\end{figure}

The number of inspected models, standard deviation of each source parameter, and Kagan angles are shown in the table 4.
\begin{figure}[htb!]
\centering
\includegraphics[width=16cm]{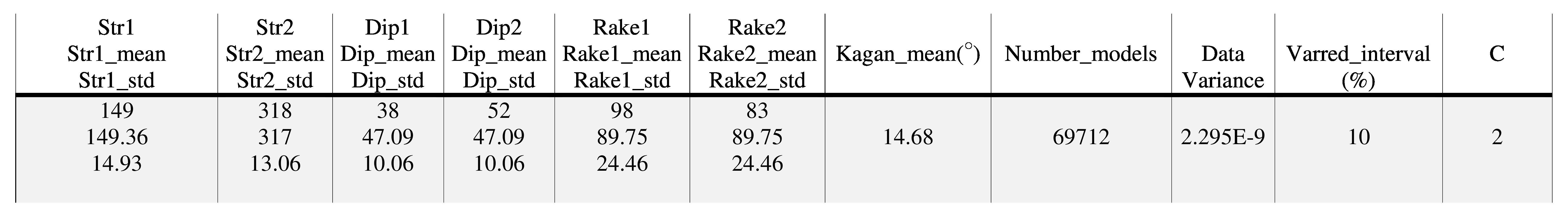}
\caption*{\small {Table 4: Average focal parameters and relevant standard deviation are shown for all the models in vicinity of the optimum solution.}}
\end{figure}

Using the approach mentioned above, theoretical error estimation of other 10 aftershocks are retrieved, and consequently their inverted solutions are selected, as is shown in the table 5.
\section{Conclusion}

With the point source inversion results of 21 events with $Mw\,\geq\,4.5$, we intend to reach a unified model describing the rupture process induced by this seismic sequence. Considering the dominant effect of error estimation of inverted depths of the events, centroid depths of the events are selected for the final conclusion, instead of relocation results. Also for some important events such as Qeshm earthquake (2005, Mw 5.8), Fin earthquake (2006, Mw 5.7), and Bam earthquake (2003, Mw 6.6), a distinctive difference between relocated and centroid depth is clearly seen, thus our approach seems reasonable. Considering a cross section with 60 degree azimuth (perpendicular to the probable causative fault), centroid depths and epicenters of 11 events with variance reduction greater than 55 percent are shown in the figure 20.
\begin{figure}[H]
\centering
\includegraphics[width=8cm]{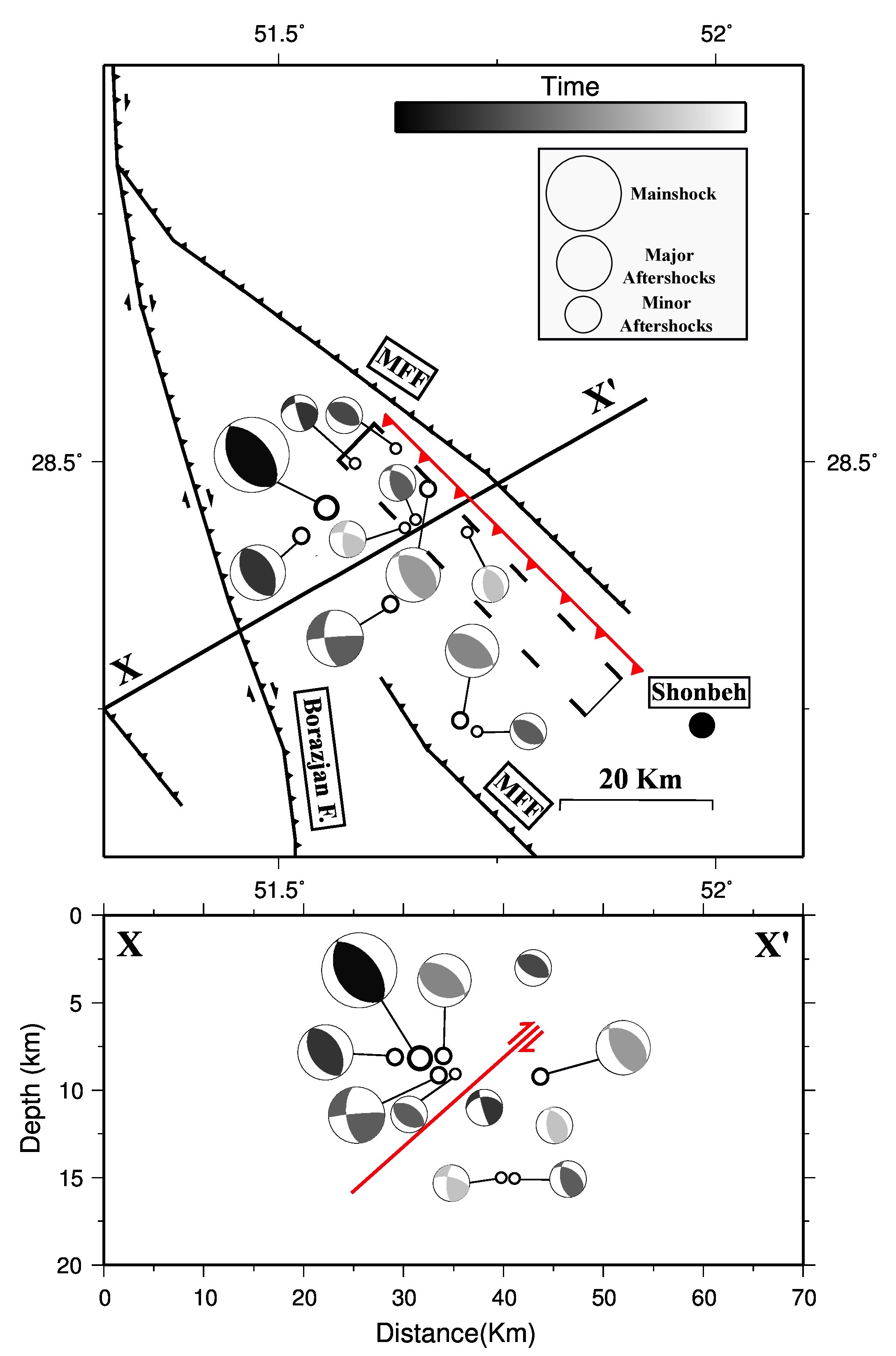}
\caption*{\small {Figure 20:  11 focal mechanisms retrieved for the events with at least $55\%$ correlation between observed and modeled waveforms. Focal spheres are colored based on the time of occurrence relative to the main shock. Reverse motion with minor left-lateral strike-slip component is considered as the dominant slip vector induced into the region of impact with a southward dip for the principle fault plane. Rupture length is estimated 40 km with 4 km width as the epicentral rupture region.}}
\end{figure}

Looking at the figure, all the events are colored based on their initiation time measured with respect to the main shock. Consequently, the migration of rupture from northwest to the southeast of the fault plane can be distinguished. Also examining the model of displacement data recorded by satellite (INSAR) during the two weeks after the main shock, we expect to have bigger shocks in the southeastern part of the rupture plane. Noting the lack of sufficient seismic energy in accord with the recorded displacements within this part, aseismic activity can be modelled for the dominant displacement in the southeastern part of the fault plane. All the inverted centroids imply southward dip with depths in the range of 8 – 12 km, suggesting 4 km width of epicentral region of the events. Also we propose a 40 km long region of impact for the sequence. The dominant rupture geometry is reverse with no explicit observable sign of the rupture on the ground. In result, this sequence may have included somehow the hidden shear sources. Based on the CMT solutions, depth range of the rupture is limited to 8-12 km implying that the dominant seismic energy is confined to the uppermost part of the bedrock and the lowermost part of the sedimentary layer. Also the sedimentary layer depth is proposed to be 12 km for this region (Oveisi et al. 2007; Oveisi et al. 2009). Noting the rupture depth, Hormoz salt unit can be responsible for the stoppage of the rupture distribution in depth. In addition, considering the slip vector of the biggest aftershock and some other bigger aftershocks, a minor left-lateral strike-slip component may be considered as the rupture mechanism.\\
\newline
\textbf{Acknowledgement}
We benifited from thoughtful suggestions and fruitful discussions with Dr. Oveisi, as he is aptly familiar with the geology and seismotectonics of the region.

\clearpage
\begin{sidewaysfigure}
    \centering
    \includegraphics[width=22cm]{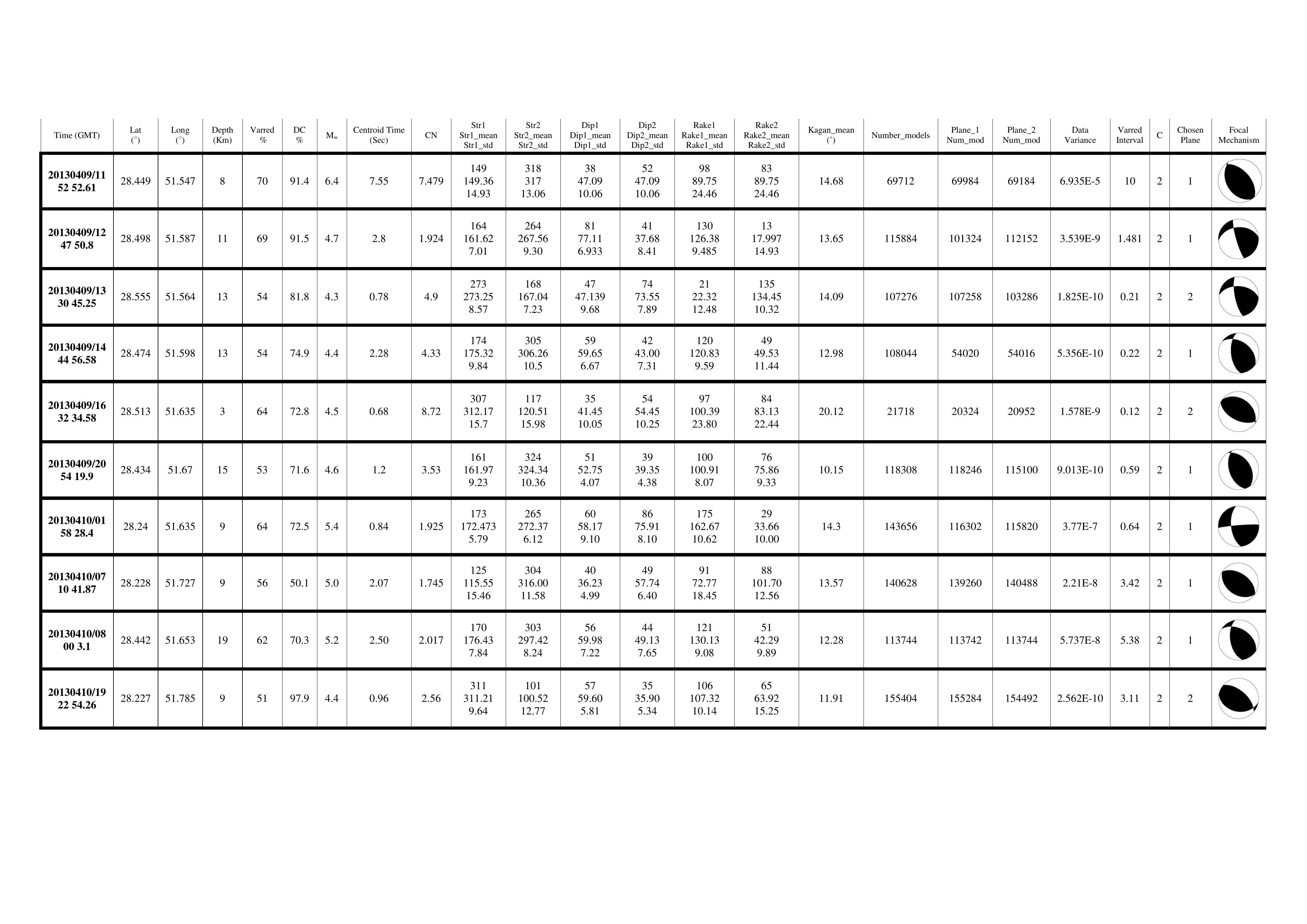}
    \caption*{\small {Table 5: Theoretical error estimation is shown for the focal mechanism of 10 events with the optimum final solution for each one. In this table, we try to choose the principle plane of incidence based on the result of theoretical error estimation of point-source solutions of the events.}}
\end{sidewaysfigure}

\end{document}